\documentclass[12pt]{article}
\pdfoutput=1

\usepackage{putex}
\usepackage{graphicx}
\usepackage{caption}
\usepackage{amsmath}
\usepackage{array}
\usepackage{subcaption}
\usepackage{epstopdf}
\usepackage{enumerate}
\usepackage{cite}
\usepackage{youngtab}
\usepackage{tensor}
\usepackage{slashed}
\usepackage[aligntableaux=center]{ytableau}
\usepackage[utf8]{inputenc}
\usepackage{rotating}
\usepackage[
      colorlinks=true,
      linkcolor=blue,
      urlcolor=blue,
      filecolor=black,
      citecolor=red,
      ]{hyperref}

\newcommand{\abs}[1]{\left\lvert #1 \right\rvert}

\newcommand {\be} {\begin {equation}}
\newcommand {\ee} {\end {equation}}

\newcommand {\bes} {\begin {equation*}}
\newcommand {\ees} {\end {equation*}}

\newcommand{\es}[2] {\begin{equation} \label{#1} \begin{split} #2 \end{split} \end{equation}}

\newcommand{\Z}{\mathbb{Z}}

\newcommand{\C}{\mathbb{C}}

\newcommand{\beq}{\begin{equation}}
\newcommand{\eeq}{\end{equation}}

\def\ie{\begin{equation}\begin{aligned}}
\def\fe{\end{aligned}\end{equation}}

\newcommand{\A}{{\alpha}}
\newcommand{\B}{{\beta}}
\newcommand{\D}{{\delta}}

\numberwithin{equation}{section}

\newcommand\Tstrut{\rule{0pt}{2.3ex}}       
\newcommand\Bstrut{\rule[-1.3ex]{0pt}{0pt}} 
\newcommand\TBstrut{\Tstrut\Bstrut}         

\def\<{\langle}
\def\>{\rangle}

\begin{document}

\preprint{PUPT-2556}

\institution{PU}{Joseph Henry Laboratories, Princeton University, Princeton, NJ 08544, USA}
\institution{HU}{Jefferson Physical Laboratory, Harvard University, Cambridge, MA 02138, USA}

\title{
The M-Theory S-Matrix From ABJM:  
\\
Beyond 11D Supergravity}

\authors{Shai M.~Chester,\worksat{\PU} Silviu S.~Pufu,\worksat{\PU} and Xi Yin\worksat{\HU}}

\abstract{
We show that by studying the flat spacetime limit of the Mellin amplitude associated with the four-point correlation function of scalar operators in the stress tensor multiplet of ABJM theory, one can produce the momentum expansion of the M-theory four-graviton S-matrix elements. Using CFT data previously obtained from the supersymmetric localization method, we carry out this procedure explicitly to the second nontrivial order in the momentum expansion, and recover precisely the known $R^4$ contribution to the scattering amplitude of super-gravitons in M-theory in eleven dimensions.
}
\date{}

\maketitle

\tableofcontents

\section{Introduction}

M-theory is a quantum theory of interacting super-gravitons in 11 dimensions with no dimensionless coupling constant \cite{Witten:1995ex}. While some of its dynamics can be understood through a combination of its relation to superstring theories via compactification and the fact that certain observables are protected by supersymmetry \cite{Horava:1996ma, Green:1997as, Russo:1997mk, Green:2005ba}, there has not been a systematic way to produce, for instance, the small momentum expansion of the graviton S-matrix in 11D Minkowskian spacetime. Neither has there been much understanding of the particle spectrum of M-theory, or lack thereof, beyond super-gravitons. 

Holographic dualities provide a window into M-theory through a dual quantum field theory. There are three important examples of such holographic duals: the Banks-Fischler-Shenker-Susskind (BFSS) matrix quantum mechanics \cite{Banks:1996vh, Balasubramanian:1997kd, Itzhaki:1998dd}, the 3-dimensional $U(N)_k\times U(N)_{-k}$ Chern-Simons-matter theory of Aharony, Bergman, Jafferis, and Maldacena (ABJM) \cite{Aharony:2008ug} in the large $N$, fixed $k$ limit,\footnote{A generalization of ABJM theory is given by $U(N)_k \times U(M)_{-k}$ Chern-Simons matter theories discussed by Aharony, Bergman, and Jafferis (ABJ) \cite{Aharony:2008gk}.  These theories are also dual to $AdS_4 \times S^7 / \Z_k$ in the limit when $N-M$ and $k$ are held fixed and $N$ is taken to infinity.} and the 6-dimensional $(2,0)$ superconformal field theories \cite{Maldacena:1997re}. The ABJM theory, dual to M-theory on $AdS_4\times S^7 / \Z_k$, is arguably the easiest to understand because it has a Lagrangian description (unlike the 6D $(2,0)$ theory) and because it has maximal superconformal symmetry when $k=1$ or $2$ (unlike BFSS matrix theory).\footnote{The same is true of the ABJ theories mentioned in the previous footnote.  Due to various dualities, the only maximally supersymmetric ABJ theory that is not dual to the $k=1$ or $k=2$ ABJM theories is the $U(N)_2 \times U(N)_{-2}$ theory.  We will not discuss this theory explicitly in this paper, although everything that we will say about the $k=2$ ABJM theory will also apply to the $U(N)_2 \times U(N)_{-2}$ ABJ theory.}  The existence of a Lagrangian description for ABJM theory allows for powerful exact results  (see, for instance, \cite{Drukker:2010nc,Kim:2009wb,Agmon:2017xes}) derived using supersymmetric localization methods (see \cite{Pestun:2016zxk} for a collection of review articles and for references), which, in recent years, have given rise to a number of precision tests of the AdS$_4$/CFT$_3$ duality:  for example, Refs.~\cite{Drukker:2010nc,Herzog:2010hf,Marino:2011eh} matched the large $N$ limit of the $S^3$ free energy of ABJM theory, computed using the supersymmetric localization results of \cite{Kapustin:2009kz}, to the same quantity computed using 11D supergravity, thus providing an impressive check of AdS/CFT at leading order in large $N$.

It has been long anticipated that the AdS/CFT correspondence  allows for extracting the full S-matrix of gravitons in the flat spacetime limit from correlation functions of the CFT \cite{Polchinski:1999ry, Susskind:1998vk, Giddings:1999jq, Penedones:2010ue, Fitzpatrick:2011jn, Fitzpatrick:2011hu}.  In practice, this approach has been hardly tractable. Recently the 4-graviton S-matrix of tree level supergravity in $AdS_4\times S^7$ has been computed in Mellin space \cite{Zhou:2017zaw} (see also \cite{Rastelli:2016nze,Rastelli:2017ymc,Rastelli:2017udc} for similar computations in $AdS_5$ and $AdS_7$) and matched \cite{Chester:2018lbz} with the leading result of the large $c_T$ expansion\footnote{$c_T$ is the coefficient of the two-point function of the canonically-normalized stress-energy tensor, as defined in Section \ref{extractcft}.  It  scales like $N^{3\over 2}$ in the large $N$, fixed $k$ limit of ABJM theory.  We prefer to think about the expansion in $1/c_T$ rather than $1/N$, because the former is what is more closely related to the expansion in Newton's constant in the flat space limit.  Note that the correlator in question is not analytic in $1/c_T$, as fractional powers and logarithmic dependence will appear in the expansion.} of the stress-energy tensor 4-point function in ABJM theory \cite{Chester:2014fya,Chester:2014mea,Agmon:2017xes}. Naturally one may wish to extend this agreement to higher orders in the large $c_T$ expansion, which amounts to going beyond supergravity in the bulk. 

In this paper, we will outline a strategy for uncovering the small momentum expansion of the 4-graviton S-matrix in M-theory from the CFT data.  At low orders in the momentum expansion, beyond the tree level terms, the S-matrix elements have local terms such as the (supersymmetrized) $R^4$ vertex, and nonlocal terms that are determined by lower order terms through unitarity cuts.  These nonlocal terms are what we loosely refer to as ``loop amplitudes" in M-theory.  Concretely, the S-matrix element involving 4 super-gravitons is constrained by supersymmetry Ward identities \cite{Dixon:1996wi,Elvang:2015rqa} to be of the form
\es{A11D}{
{\cal A} = f(s, t) {\cal A}_{\rm SG, tree} \,,
 }
where ${\cal A}_{\rm SG, tree}$ is the tree level scattering amplitude in 2-derivative supergravity, and $f$ is a symmetric function of the Mandelstam invariants $s$, $t$, and $u=-s-t$.  The tree-level supergravity scattering amplitude ${\cal A}_{\rm SG, tree}$ carries dependence on the polarization as well as the type of particles in the super-graviton multiplet.  The function $f$ admits a small momentum expansion, or equivalently, an expansion in the 11D Planck length $\ell_{11}$, of the form
 \es{fExpansion}{
  f(s, t) = 1 + \ell_{11}^6 f_{R^4}(s, t) 
   + \ell_{11}^9 f_\text{1-loop}(s, t)  + \ell_{11}^{12} f_{D^6 R^4}(s, t) + \ell_{11}^{14} f_{D^8 R^4}(s, t)
   + \cdots \,.
 }
Here, $f_{D^{2n}R^4}$ refers to a local term which is a degree $n+3$ symmetric polynomial in $s, t, u$, whereas the loop terms are not analytic at zero momentum. In particular, $f_{R^4}$, $f_{\rm 1-loop}$, and $f_{D^6R^4}$ are known exactly \cite{Green:1997as, Russo:1997mk, Green:2005ba}, as they are protected by supersymmetry and can be determined by perturbative calculations either in type II string theory or in 11D supergravity \cite{Green:1998by, Pioline:2015yea}.  For instance,
 \es{fR4}{
  f_{R^4} (s, t, u) = \frac{ s t u}{3 \cdot 2^7} , ~~~~~ f_{D^6R^4}(s,t,u) = {(stu)^2\over 15\cdot 2^{15}}.
 }
Note that a term of the form $\ell_{11}^{10} f_{D^4R^4}(s,t)$ in (\ref{fExpansion}) would be allowed by the supersymmetric Ward identity, but it is known to be absent by comparison with type II string scattering amplitudes and supersymmetry renormalization properties. The term $\ell_{11}^{14} f_{D^8R^4}(s,t)$, on the other hand, is not protected by supersymmetry, and its existence is not known to the best of our knowledge (although it was conjectured to be absent in \cite{Russo:1997mk}). 

As mentioned above, our goal here is to reproduce the expansion \eqref{fExpansion} by taking the flat space limit of the CFT correlators.  We will carry out this strategy to the first nontrivial order beyond two-derivative supergravity, and produce the $R^4$ effective coupling of M-theory from the large $c_T$ expansion of a known BPS OPE coefficient in ABJM theory, in the following steps:

\bigskip

\begin{enumerate}[(1)]

\item We focus on the 4-point function of dimension $\Delta=1$ scalar primaries $S_{IJ}$ in the stress tensor supermultiplet that transform in the ${\bf 35}_c$ representation of $\mathfrak{so}(8)$ R-symmetry, in ABJM theory with $k=1$ or $2$. Its Mellin transform, to be defined in Section~\ref{stresscorr}, admits a large $c_T$ expansion of the form
\ie
\label{mellinexp}
M(s,t;\sigma,\tau) = c_T^{-1} M_{\rm tree} + c_T^{-{5\over 3}} M_{R^4} + c_T^{-2} M_{\rm 1-loop} + \cdots \,.
\fe
Here, $s, t$ are Mellin space kinetic variables (not to be confused with the Mandelstam invariants), and $\sigma,\tau$ are $\mathfrak{so}(8)$ invariants that will be defined in Section~\ref{stresscorr}. $M_\text{tree}$ represents the tree-level supergravity contribution, recently computed in \cite{Zhou:2017zaw}. $M_{R^4}$ is a polynomial expression in $s, t$, whose large $s, t$ limit will be related to the 4-graviton vertex that corresponds to the $R^4$ effective coupling in flat spacetime. $M_{\rm 1-loop}$ is the 1-loop supergravity contribution in $AdS_4\times S^7$, which is free of logarithmic divergences. The higher order terms in the expansion may involve logarithmic dependence on $c_T$, as we will discuss later.

\item At each order in the $1/c_T$ expansion, the Mellin amplitude is subject to the ${\cal N}=8$ superconformal Ward identity. If the amplitude is a polynomial in $s, t$ of known maximal degree, e.g.~$M_{R^4}$ is a degree 4 polynomial expression, then the Ward identity allows for finitely many solutions, thereby constraining the Mellin amplitude at this order in terms of finitely many unknown coefficients. Some linear combinations of these coefficients will be related to flat space amplitudes through the large $s, t$ limit. The ``loop amplitudes"  will be determined by lower order terms in the $1/c_T$ expansion up to residual polynomial terms.  Note that the loop Mellin amplitudes involve sums over poles that correspond to multi-trace operators in the OPE, and in the flat space limit the poles turn into branch cuts.

\item Transforming the Mellin amplitude back to the correlation function, one would recover from (\ref{mellinexp}) the $1/c_T$ expansion of the OPE coefficients as well as the scaling dimensions of various unprotected superconformal primaries. Some of these OPE coefficients, namely those of certain $1/2$-BPS and $1/4$-BPS multiplets, are known exactly as a function of $c_T$ from supersymmetric localization computations \cite{Chester:2014fya,Chester:2014mea,Agmon:2017xes}. Other OPE coefficients, as well as the scaling dimension of long multiplets, are not known exactly but can be constrained by conformal bootstrap bounds.

\item We will see that the OPE coefficient of the $1/4$-BPS $(B,2)$ multiplet, expanded to order $c_T^{-{5\over 3}}$, determines the coefficient of $M_{R^4}$ in \eqref{mellinexp}. Taking its large $s,t$ limit then determines the $R^4$ effective coupling of M-theory in flat spacetime.\footnote{One may contemplate, in principle, a more powerful approach for determining the couplings in the M-theory effective action, as follows.  In principle, 11d SUSY determines the supersymmetric completion of the $D^{2k} R^4$ terms (perhaps up to a few coefficients).  One can then reduce the 11d action on $S^7$ to obtain an effective action in $AdS_4$, which can then be used to calculate the CFT data via Witten diagrams.  In practice, none of these steps are currently achievable without a tremendous effort.  We thank Ofer Aharony for this comment.}   Our result is in perfect agreement with the known $R^4$ coefficient in \eqref{fR4}, previously derived by combining toroidal compactification of M-theory, comparison to perturbative type II string amplitudes, and protection by supersymmetry.

\end{enumerate}

It is worth noting that previously, in the AdS/CFT context, the $R^4$ coupling of M-theory has been probed through the study of conformal anomaly of the 6D $(2,0)$ theory \cite{Tseytlin:2000sf}. In this approach, one makes use of the bulk Lagrangian, including $R^4$ coupling as well as other terms related by supersymmetry. However, it is difficult to justify whether one has accounted for all the relevant terms in the effective Lagrangian, which is further subject to the ambiguity of field redefinitions. In contrast, our strategy produces from CFT data terms in the flat space S-matrix element, it is not subject to complications of the bulk effective Lagrangian, and all supersymmetries are manifest \cite{Bianchi:2008pu, Elvang:2010jv}.

A related comment concerns the structure of the derivative expansion of M-theory in 11D flat spacetime. Absent a dimensionless coupling constant, one could either speak of a Wilsonian effective Lagrangian, which is subject to the ambiguity of a floating cutoff scheme, or the 1PI/quantum effective Lagrangian, which amounts to a generating functional for the graviton S-matrix and is nonlocal. It is accidental, thanks to supersymmetry, that low order terms in the derivative expansion of the 1PI effective Lagrangian of M-theory can be separated into local terms, such as $t_8 t_8 R^4$, and nonlocal terms that correspond to loop amplitudes. This distinction ceases to exist starting at 20-th derivative order, where the supergravity 2-loop amplitude has a logarithmic divergence that is cut off at the Planck scale and mixes with a local term of the schematic form $D^{12} R^4$ \cite{Bern:1998ug, Green:1999pu}.  As mentioned above, it is clearer to phrase all of this directly in the language of the graviton S-matrix, and its expansion at small momenta as given in Eq.~\eqref{fExpansion}.

Finally, we should note that the idea that a large $N$ CFT has a finite number of solutions to the conformal Ward identities at each order in $N$ was first stated in \cite{Heemskerk:2009pn}. In subsequent work \cite{Alday:2014tsa,Aprile:2017bgs}, this idea was generalized to maximally supersymmetric SCFTs in 4D and 6D, respectively, where the superconformal Ward identities further constrain the number of solutions. In 4D, \cite{Goncalves:2014ffa} related the flat space limit of the Mellin amplitude to the S-matrix of type IIB string theory in 10D, but a precise reconstruction of the 10D S-matrix was not possible because of a lack of known CFT data that can fix the undetermined parameter in the CFT 4-point function.   In the present work, we provide the first application of these ideas to 3D, and, as mentioned above, we can further recover the $R^4$ term in 11D from the CFT correlators by making use of nontrivial CFT data that can be computed using supersymmetric localization.\footnote{In 4D and 6D there exists a protected part of the 4-point function of the $1/2$-BPS scalar in the stress tensor multiplet that can be computed exactly \cite{Beem:2013sza,Beem:2014kka}.  This sector, however, is completely fixed at order $1/c_T$, i.e.~supergravity, for the stress tensor four point function. }

The rest of this paper is organized as follows.  In Section~\ref{stresscorr} we start with a brief review of the properties of the four-point function of the scalar operators $S_{IJ}$ in the stress tensor multiplet of a local ${\cal N}= 8$ SCFT\@.  For ABJM theory, we also summarize the known exact results on OPE coefficients derived using supersymmetric localization.  In Section~\ref{HOLOGRAPHIC} we use the superconformal Ward identity as well as the asymptotic growth conditions on the Mellin amplitude in order to determine, up to a few constants, the Mellin amplitude order by order in $1/c_T$ in the case of the M2-brane theory.  In Section~\ref{extractcft} we explain how to extract various scaling dimensions and OPE coefficients from the Mellin amplitude constructed in Section~\ref{HOLOGRAPHIC}, and show how to reproduce the known correction to the supergravity scattering amplitude of four super-gravitons in 11D\@.  Lastly, we end in Section~\ref{CONCLUSION} with a brief summary as well as a discussion of future directions.

\section{Four point function of stress-tensor}
\label{stresscorr}

Let us start by reviewing some general facts about 3D ${\cal N} = 8$ local SCFTs and of the constraints imposed by the $\mathfrak{osp}(8|4)$ algebra.  The discussion in this section is quite general;  it does not rely on the Lagrangian of a particular 3D ${\cal N} = 8$ SCFT, nor does it assume such a theory has a holographic dual.

As mentioned in the Introduction, all ${\cal N} =8$ local SCFTs have a stress energy tensor which belongs to the same half-BPS multiplet as a scalar operator of scaling dimension $\Delta = 1$ transforming, by convention, in the ${\bf 35}_c$ representation of the $\mathfrak{so}(8)$ R-symmetry.  (We choose the eight supercharges also by convention to transform in the ${\bf 8}_v$.)  Since analyzing correlation functions of scalar operators is easier than analyzing correlators of the stress tensor, we will focus on these scalar operators.

We can view the ${\bf 35}_c$ representation as the rank-two symmetric traceless product of the ${\bf 8}_c$, and so our scalar operators are traceless symmetric tensors $S_{IJ}(\vec{x})$, where $I, J = 1, \ldots 8$ are ${\bf 8}_c$ indices.  In order to not carry around the ${SO}(8)$ indices, it is convenient to contract them with an auxiliary polarization vector $Y^I$ that is constrained to be null $Y \cdot Y \equiv \sum_{I=1}^8 (Y^I)^2 = 0$, thus defining 
 \es{ODef}{
  S(\vec{x},Y) \equiv S_{IJ}(\vec{x}) Y^I Y^J \,.
 }
Conformal symmetry and $\mathfrak{so}(8)$ symmetry imply that the four point function of  $S(\vec{x}, Y)$ takes the form 
 \es{FourPointO}{
  \left\langle S(\vec{x}_1,Y_1) S(\vec{x}_2,Y_2)
   S(\vec{x}_3,Y_3) S(\vec{x}_4,Y_4) \right\rangle&=\frac{( Y_1\cdot Y_2 )^2  ( Y_3\cdot Y_4 )^2 }
     {\abs{\vec{x}_{12}}^{2}\abs{\vec{x}_{34}}^{2} }\mathcal{G}(U, V;\sigma,\tau) \,,
 }
where $U$ and $V$ are conformally-invariant cross ratios, and $\sigma$ and $\tau$ are $\mathfrak{so}(8)$ invariants formed out of the polarizations
 \es{uvsigmatauDefs}{
  U \equiv \frac{\vec{x}_{12}^2 \vec{x}_{34}^2}{\vec{x}_{13}^2 \vec{x}_{24}^2} \,, \qquad
   V \equiv \frac{\vec{x}_{14}^2 \vec{x}_{23}^2}{\vec{x}_{13}^2 \vec{x}_{24}^2}  \,, \qquad
   \sigma\equiv\frac{(Y_1\cdot Y_3)(Y_2\cdot Y_4)}{(Y_1\cdot Y_2)(Y_3\cdot Y_4)}\,,\qquad \tau\equiv\frac{(Y_1\cdot Y_4)(Y_2\cdot Y_3)}{(Y_1\cdot Y_2)(Y_3\cdot Y_4)} \,.
 }
Because \eqref{FourPointO} is a quadratic polynomial in each $Y_i$ separately, the quantity $\mathcal{G}(U,V;\sigma,\tau)$ is a quadratic function of $\sigma$ and $\tau$. 
 
By performing the OPE between the first two and last two operators in \eqref{FourPointO}, one can decompose $ \mathcal{G}(U,V;\sigma,\tau)$ into superconformal blocks $\mathfrak{G}_{\mathcal{M}}$, 
\es{SBDecomp}{
     \mathcal{G}(U,V;\sigma,\tau)=\sum_{\mathcal{M}\in\mathfrak{osp}(8|4)}\lambda^2_\mathcal{M}\mathfrak{G}_\mathcal{M}(U,V;\sigma,\tau)\,,
}
where ${\cal M}$ runs over all $\mathfrak{osp}(8|4)$ multiplets appearing in the $S \times S$ OPE, and the $\lambda^2_{\mathcal{M}}$ are the squared OPE coefficients for each such supermultiplet $\mathcal{M}$.  In Table \ref{opemult}, we list the multiplets $\mathcal{M}$ that appear in the OPE $S\times S$, the dimension, spin, and $\mathfrak{so}(8)$ representation of their primaries, along with the possible values of their Lorentz spins. In our notation, the $(B,+)$ $[0020]$ multiplet with $(\Delta,j)=(1,0)$ is the stress tensor multiplet itself. Unless otherwise noted, the $(B,+)$ $[0040]$ multiplet in the $S\times S$ OPE will be simply be referred to as the ``$(B,+)$ multiplet".\footnote{Note that for the interacting ABJ(M) theories, there are two degenerate $(B,+)$ $[0040]$ multiplets, corresponding to a single trace (a super-graviton KK mode) as well as a double trace operator. The one that enters the $S\times S$ OPE is a specific linear combination of the two, and is dominated by the double trace operator in the large $N$ limit. All other multiplets appearing in Table \ref{opemult} involve multi-trace operators in the large $N$ limit, barring the possibility of stable Planckian particles that could show up as single trace long multiplets, whose scaling dimensions would be of order $N^{1\over 6}$.} Its OPE coefficient will be denoted $\lambda_{(B,+)}$. Likewise the OPE coefficient of the $(B,2)$ multiplet will be denoted $\lambda_{(B,2)}$. The semi-short multiplets are denoted $(A,2)_j$ and $(A,+)_j$ where $j$ is the spin. 
 
Perturbatively in $1/N$, the long multiplets that appear in the OPE will be denoted $(A,0)_{n,j,q}$, where $n=0,1,\dots$ labels the leading order twist $2n+2$, and $q=0\,,\dots n$ is an index that labels distinct operators with the same leading order quantum numbers. Subleading corrections in $1/N$ will lift the degeneracy among these long multiplets. In the $n=0$ case, we will omit the label $q(=0)$ and denote the multiplet by $(A,0)_{0,j}$. 
 
\begin{table}
\centering
\begin{tabular}{|c|c|r|c|c|}
\hline
Type    & $(\Delta,j)$     & $\mathfrak{so}(8)$ irrep  & spin $j$ & BPS\\
\hline
$(B,+)$ &  $(2,0)$         & ${\bf 294}_c = [0040]$   & $0$ & 1/2 \\ 
$(B,2)$ &  $(2,0)$         & ${\bf 300} = [0200]$  & $0$ &1/4  \\
$(B,+)$ &  $(1,0)$         & ${\bf 35}_c = [0020]$ & $0$ &1/2 \\
$(A,+)$ &  $(j+2,j)$       & ${\bf 35}_c = [0020]$ &even &1/4 \\
$(A,2)$ &  $(j+2,j)$       & ${\bf 28} = [0100]$ & odd &1/8 \\
$(A,0)$ &  $\Delta\ge j+1$ & ${\bf 1} = [0000]$ & even& Long \\
\hline
\end{tabular}
\caption{The possible superconformal multiplets in the $S\times S$ OPE\@.  The $\mathfrak{so}(3, 2) \oplus \mathfrak{so}(8)$ quantum numbers are those of the superconformal primary in each multiplet.}
\label{opemult}
\end{table}
 
The superconformal block $\mathfrak{G}_{\mathcal{M}}$ corresponding to each of the multiplets listed in Table~\ref{opemult} receives contributions from conformal primaries with different spins $j'$, scaling dimensions $\Delta'$, and $\mathfrak{so}(8)$ irreps $[0\, (a-b)\, (2b)\, 0]$ for $a=0,1,2$ and $b=0,\dots, a$ that appear in the tensor product $[0020]\otimes [0020]$. The superconformal block can thus be written as a linear combination of the conformal blocks $G_{\Delta',j'}$ corresponding to the conformal primaries in $\mathcal{M}$ as
 \es{GExpansion}{
  {\mathfrak G}_{\mathcal{M}}(U, V; \sigma, \tau) = \sum_{a=0}^2 \sum_{b = 0}^a Y_{ab}(\sigma, \tau)  \sum_{(\Delta',j')\in\mathcal{M}} A^\mathcal{M}_{ab \Delta' j'}(\Delta, j) G_{\Delta',j'}(U,V) \,.
 }  
Here, the quadratic polynomials $Y_{ab}(\sigma, \tau)$ are eigenfunctions of the $\mathfrak{so}(8)$ Casimir corresponding to the various irreducible $\mathfrak{so}(8)$ representations appearing in the product ${\bf 35}_c \otimes {\bf 35}_c$, and are given by \cite{Dolan:2003hv,Nirschl:2004pa}
 \es{polyns}{
   {\bf 1} = [0000]: \qquad Y_{00}(\sigma, \tau) &= 1 \,, \\
    {\bf 28} = [0100]: \qquad Y_{10}(\sigma, \tau) &= \sigma - \tau \,, \\
   {\bf 35}_c = [0020]: \qquad  Y_{11}(\sigma, \tau) &= \sigma + \tau -\frac{1}{4} \,, \\
   {\bf 300} = [0200]: \qquad Y_{20}(\sigma, \tau) &= \sigma^2 + \tau^2 - 2\sigma\tau - \frac{1}{3}(\sigma + \tau) + \frac{1}{21} \,,\\
    {\bf 567}_c = [0120]: \qquad Y_{21}(\sigma, \tau) &= \sigma^2 - \tau^2 - \frac{2}{5}(\sigma - \tau) \,,\\
    {\bf 294}_c = [0040]: \qquad Y_{22}(\sigma, \tau) &= \sigma^2 + \tau^2 + 4\sigma\tau - \frac{2}{3}(\sigma+\tau) + \frac{1}{15} \,.
 }
The $A^\mathcal{M}_{ab\Delta'j'}(\Delta, j)$ are rational function of $\Delta$ and $j$. For the list of conformal primaries that appear in each $\mathfrak{G}_\mathcal{M}$ as well as the explicit coefficients $A^\mathcal{M}$, see \cite{Chester:2014fya}.

\subsection{The Mellin amplitude}
\label{MELLIN}

Any 4-point function of scalar operators can be equivalently expressed in Mellin space.  We will find it useful to separate out the disconnected piece of the correlator, which in a convenient normalization for $S_{IJ}$ takes the form
 \es{GDisc}{
  {\cal G}_\text{disc}(U, V; \sigma, \tau) 
   = 1 + U \sigma^2 + \frac{U}{V} \tau^2 \,,
 } 
and then define the Mellin transform just for the connected part $\mathcal{G}_\text{conn} \equiv {\cal G} - \mathcal{G}_\text{disc}$:
\es{mellinDef}{
\mathcal{G}_\text{conn}(U,V;\sigma,\tau)=\int_{-i\infty}^{i\infty}\frac{ds\, dt}{(4\pi i)^2} U^{\frac s2}V^{\frac t2-\Delta}M(s,t;\sigma,\tau)\Gamma^2\left[\Delta-\frac s2\right]\Gamma^2\left[\Delta-\frac t2\right]\Gamma^2\left[\Delta-\frac u2\right]\,.
}
Here, the Mellin space variables $s$, $t$, and $u$ satisfy the constraint $s+t+u=4\Delta$, and recall that for our 4-point function $\Delta=1$. The two integration contours run parallel to the imaginary axis, such that all poles of the Gamma functions are on one side or the other of the contour.

\subsection{Localization results for short operators in ABJ(M)}

The OPE coefficients of all short $B$-type operators in Table \ref{opemult} have been computed in \cite{Agmon:2017xes} for all 3D $\mathcal{N}=8$ theories, making use of a topological subsector of these theories studied in \cite{Chester:2014mea,Beem:2016cbd,Dedushenko:2016jxl}.\footnote{The generalization of the methods of \cite{Dedushenko:2017avn} to non-Abelian theories would allow for a more direct computation of these OPE coefficients, without relying on the approach of \cite{Agmon:2017xes}.} For ABJM with gauge group $U(N)_k\times U(N)_{-k}$ and ABJ with $U(N+1)_2\times U(N)_{-2}$, the result is known perturbatively to all orders in the large $N$ limit, or exactly at small values of $N$. To compare to gravity, it is more convenient to reorganize the large $N$ expansion in terms of an expansion in the inverse of $c_T$, the coefficient of the canonically normalized stress tensor two-point function
 \es{TmnCorr}{
  \langle T_{\mu\nu}(x) T_{\rho \sigma}(0) \rangle =  \frac{c_T}{64}\left( P_{\mu\rho} P_{\nu\sigma} + P_{\nu \rho} P_{\mu \sigma} - P_{\mu\nu} P_{\rho\sigma} \right) \frac{1}{16 \pi^2 x^2} \,,
 }
where $P_{\mu\nu} \equiv \eta_{\mu\nu} \nabla^2 - \partial_\mu \partial_\nu$. Our convention in \eqref{TmnCorr} is such that $c_T=1$ for a 3D (non-supersymmetric) theory of a real massless scalar or of a Majorana fermion. In this convention $c_T$ is related to the stress tensor OPE coefficient as
\es{stresstocT}{
c_T=\frac{256}{\lambda^2_S}\,.
}
 For $U(N)_k\times U(N)_{-k}$ ABJM theory, the large $N$ expansion of $c_T$ is
\es{cT}{
c_T=\frac{64}{3\pi}\sqrt{2k}N^{3/2}+\frac{32\sqrt{2}}{\pi\sqrt{k}}N^{1/2}+O(N^0)\,.
}
The large $N$ expansion of the OPE coefficients of the $(B,+)$ and $(B,2)$ operators can then be rewritten as an expansion in large $c_T$ to all orders. The first few terms are\footnote{It can be argued that the $1/c_T$ expansion of these OPE coefficients is perturbatively the same for the $U(N+1)_2\times U(N)_{-2}$ ABJ theory and the $U(N)_2\times U(N)_{-2}$ ABJM theory.  From the M-theory point of view, the two theories differ by a torsion flux, i.e.~a discrete holonomy of the 3-form field on a torsion 3-cycle of $S^7 / \Z_2$ \cite{Aharony:2008gk}.  This torsion flux affects the CFT data only through non-perturbative effects.\label{FootnoteNonPert}}
\es{exact}{
\lambda^2_{(B,2)}&=\frac{32}{3}-\left(\frac{4096}{9}-\frac{5120}{3\pi^2}\right)c_T^{-1}+{40960}\left(\frac{2}{9\pi^8k^2}\right)^{\frac13}c_T^{-\frac53}+O(c_T^{-2})\,,\\
\lambda^2_{(B,+)}&=\frac{16}{3}+\left(\frac{1024}{9}+\frac{1024}{3\pi^2}\right)c_T^{-1}+{8192}\left(\frac{2}{9\pi^8k^2}\right)^{\frac13}c_T^{-\frac53}+O(c_T^{-2})\,,\\
}
where note that the $k$-dependence begins at order $c_T^{-5/3}$, and all higher order terms appear at orders $c_T^{-\frac{2}{3}n}$ and $c_T^{-\left(\frac{2}{3}n+1\right)}$ for integer $n$. In fact, it can be argued that only one of the OPE coefficients in \eqref{exact} is independent, because the relation \cite{Chester:2014mea}
\es{1dcrossing}{
\frac{1024}{c_T}-5\lambda^2_{(B,+)}+\lambda^2_{(B,2)}+16=0
}
is a consequence of crossing symmetry and must hold exactly.

\section{The holographic four-point function}
\label{HOLOGRAPHIC}

Let us now discuss the 4-point correlator of the operators $S_{IJ}$ in the particular case of ABJM theory at CS level $k=1$ or $2$.  In this section, we will use the AdS/CFT duality to study this correlator from the bulk side of the duality, without making any reference to the ABJM Lagrangian.  We will use, however, that this theory is the low-energy theory on $N$ coincident  M2-branes placed at a $\C^4 / \Z_k$ singularity, and that perturbatively at large $N$ the back-reacted geometry is $AdS_4 \times S^7/\Z_k$.  The radius $L$ of $AdS_4$ is given by
 \es{LellpRelation}{
  \frac{L^6}{\ell_{11}^6} = \frac{N k}{8}  + O(N^0)=\left(\frac{3\pi c_T k}{2^{11}}\right)^{\frac23}+O(c_T^0) \,,
 }
where $\ell_{11}$ is the 11D Planck length \cite{Aharony:2008ug}.\footnote{In the ABJM paper \cite{Aharony:2008ug}, the radius of AdS is $L$ is denoted by $R/2$.  Eq.~(4.2) in that paper then implies $L^6 / \ell_p^6 =  \pi^2 N k / 2$.  The scattering amplitudes in the main text were written in the convention $2 \kappa_{11}^2 = (2 \pi)^5 \ell_{11}^9$ whereas the ABJM paper uses the Polchinski \cite{Polchinski:1998rr} convention $2 \kappa_{11}^2 = (2 \pi)^8 \ell_p^9$.  Thus, $\ell_p = \ell_{11} (2 \pi)^{-1/3}$, so $L^6 / \ell_{11}^6 =  N k / 8$.}   At leading order in $1/N$, the radius of $S^7 / \Z_k$ is equal to $2L$. 

Note that the subleading corrections in (\ref{LellpRelation}) depend on the precise definition of $L$ beyond the supergravity solution. This ambiguity will not be important for us, as the precise large radius expansion will be performed in $1/c_T$ rather than in $\ell_{11}/L$.

\subsection{Holographic correlator in tree level supergravity}

The main advantage of the Mellin space representation mentioned in Section~\ref{MELLIN} is that in a theory with a holographic dual one can easily write down the {\em tree level} expression for the connected part of the four-point function.  Indeed, the simplicity comes about as follows.  At tree level, the relevant Witten diagrams are contact diagrams and exchange diagrams, so 
 \es{Mtree}{
  M_\text{tree} =  M_\text{$s$-exchange} +M_\text{$t$-exchange}+M_\text{$u$-exchange}   + M_\text{contact}  \,,
 }
while the $t$- and $u$-channel exchange diagrams are related to the $s$-channel one as
\es{tandu}{
M_\text{$t$-exchange}(s,t;\sigma,\tau)&=\tau^2 M_\text{$s$-exchange}(t,s;\sigma/\tau,1/\tau)\,,\\
M_\text{$u$-exchange}(s,t;\sigma,\tau)&=\sigma^2 M_\text{$s$-exchange}(u,t;1/\sigma,\tau/\sigma)\,.
}
In Mellin space, the contact diagrams corresponding to vertices with $n$ derivatives are order $n$ polynomials in $s$, $t$, $u$.  The exchange diagrams are slightly more complicated.  An exchange diagram for a bulk field $\phi$ dual to a boundary conformal primary operator ${\cal O}$ of dimension $\Delta_{\cal O}$ and spin $\ell_{\cal O}$ has a meromorphic piece whose form is fixed up to an overall constant by the requirement that the residue at each pole agrees with the residue of the conformal block corresponding to the exchange of the operator ${\cal O}$, as well as a polynomial piece in $s$, $t$, $u$. The degree of the polynomial is given by $p_1 + p_2 -1$, where $p_1$ and $p_2$ are half the numbers of derivatives in the two vertices connecting the $\phi$ internal line to the external lines.  The meromorphic piece is independent of the vertices, and it has poles at $s = 2m + \tau_{\cal O}$, where $\tau_{\cal O} = \Delta_{\cal O} - \ell_{\cal O}$ is the twist of the conformal primary ${\cal O}$, and $m= 0, 1, 2, \ldots$.  For example, if we denote
 \es{Mexchange}{
  M_\text{$s$-exchange}^\phi = \widehat{M}_\text{$s$-exchange}^\phi + \text{(analytic)} \,,
 }
then the meromorphic pieces for various bulk fields that will be of interest to us can be taken to be:\footnote{These expressions are just rescaled versions of (3.31) of \cite{Zhou:2017zaw}.  In particular, we have
\begin{equation*}
\begin{aligned}
  \widehat{M}_\text{$s$-exchange}^\text{graviton}
   &= -\sum_{n=0}^\infty \frac{  \cos (n \pi) \Gamma(-\frac 32 - n )}{4 \sqrt{\pi} n! \Gamma(1/2 - n)^2} \frac{4n^2 - 8ns + 8n + 4s^2 + 8 s t - 20 s + 8 t^2 - 32 t + 35}{s - (2n+1)}
    = -\frac{M_\text{graviton}^\text{Zhou}}{3 \pi} \,, \\
   \widehat{M}_\text{$s$-exchange}^\text{gauge field}&=
   - \sum_{n=0}^\infty \frac{\cos (n \pi)}{\sqrt{\pi}  (1 + 2n) \Gamma(\frac 12-n) \Gamma(1 + n)} \frac{2t + s - 4}{s - (2n+1)} = -\frac{M_\text{vector}^\text{Zhou}}{ \pi} \,, \\
    \widehat{M}_\text{$s$-exchange}^\text{$\Delta = 1$ scalar}
     &=  -4 \sum_{n=0}^\infty \frac{ \cos (n \pi)}{\sqrt{\pi} n! \Gamma(\frac 12-n) } \frac{1}{s - (2n+1)}
      =-4 \frac{M_\text{scalar}^\text{Zhou}}{\pi} \,.
  \end{aligned}
  \end{equation*}
 }
\es{sChannel}{
\widehat{M}_\text{$s$-exchange}^\text{graviton}&=
  \frac{t^2 + u^2 - 6 t u + 6 (t + u) - 8}{4s(s+2)} \left( \frac{- (s+4)}{2} + \widehat{M}_\text{$s$-exchange}^\text{$\Delta = 1$ scalar} \right) 
   - \frac{(3s-4)}{8}  \,,\\
\widehat{M}_\text{$s$-exchange}^\text{gauge field}&=
 \frac{t-u}{4s} \left(-2+ \widehat{M}_\text{$s$-exchange}^\text{$\Delta = 1$ scalar} \right)  \,,\\
\widehat{M}_\text{$s$-exchange}^\text{$\Delta = 1$ scalar}&=
 \frac{ 2\Gamma \left(\frac{1-s}{2} \right)}{ \sqrt{\pi} \Gamma\left(1 - \frac{s}{2} \right)} \,.
}
In addition, we note that the contribution from any bulk field $\phi$ dual to an even-twist conformal primary must vanish:
  \es{sChannelEvenTwist}{
  \widehat{M}_\text{$s$-exchange}^\text{even twist $\phi$}(s, t)&= 0 \,,
  }
because a non-zero meromorphic piece for such an exchange would have poles at even values of $s$, and that would produce third order poles when inserted in \eqref{mellinDef}.

Going back to the situation of interest to us, i.e.~the four-point function of the $S_{IJ}$ operators in the $k=1$ ABJM theory,\footnote{The computation for the $k=2$ ABJ(M) theory is identical at leading order in the $1/c_T$ expansion.} we should think about which exchange and contact diagrams we should write down.  The scalar operators $S_{IJ}$ are dual to certain components of the 11D graviton and 3-form in the $S^7$ directions.  As is well known, the spectrum of fluctuations around $AdS_4 \times S^7$ organizes into representations of the supersymmetry algebra $\mathfrak{osp}(8|4)$ \cite{Biran:1983iy} (which is the same as the 3D ${\cal N}= 8$ superconformal algebra). As shown in Table \ref{opemult}, the $S\times S$ OPE contains two half-BPS operators: the stress tensor multiplet whose bottom component is $S$ itself, and the $(B,+)$ multiplet whose component operators all have even twist. From the discussion above, it follows that the only bulk fields that contribute a meromorphic piece in the exchange diagrams are those in the stress tensor multiplet: the scalar fields dual to $S$, the $\mathfrak{so}(8)$ gauge fields, and the graviton.\footnote{There is no bulk coupling between three scalars in the gravity multiplet, but there exists a boundary term that couples them (see for instance \cite{Freedman:2016yue}).  Therefore in the scalar exchange diagram the two intermediate points are located on the boundary.}  Consequently, $M_\text{$s$-exchange}$ is (up to an overall normalization that we will introduce later) a linear combination\footnote{In the notation of \cite{Zhou:2017zaw}, we have $\lambda_s =- 1/\pi$, $\lambda_v = -b/\pi$, and $\lambda_g = -c/(3\pi)$. }
 \es{Ms}{
   M_\text{$s$-exchange} =Y_{{\bf 35}_c}(\sigma, \tau) M_\text{$s$-exchange}^\text{$\Delta = 1$ scalar}   + b  Y_{\bf 28}(\sigma, \tau) M_\text{$s$-exchange}^\text{gauge}+ c Y_{\bf 1}(\sigma, \tau) M_\text{$s$-exchange}^\text{graviton} \,,
 } 
for some constants $b$ and $c$.  To determine the relative coefficients one can use the superconformal Ward identity (see Appendix~\ref{SUSYWARD}), which, as shown in \cite{Zhou:2017zaw}, implies $b = -4$ and $c = 1$, so
 \es{MsAgain}{
   M_\text{$s$-exchange} = Y_{{\bf 35}_c}(\sigma, \tau) M_\text{$s$-exchange}^\text{$\Delta = 1$ scalar}  -   4 Y_{\bf 28}(\sigma, \tau) M_\text{$s$-exchange}^\text{gauge}+   Y_{\bf 1}(\sigma, \tau) M_\text{$s$-exchange}^\text{graviton}   \,.
 }
Consequently, we can write a general tree-level Mellin amplitude as
 \es{MtreeAgain}{
  M_\text{tree} =  C \left[  \widehat{M}_\text{exchange} + M_\text{residual} \right] \,,
 } 
where $\widehat{M}_\text{exchange}= \widehat{M}_\text{$s$-exchange} +\widehat{M}_\text{$t$-exchange}+\widehat{M}_\text{$u$-exchange}$, $\widehat{M}_\text{$s$-exchange}$ is given by \eqref{MsAgain} with all $M$'s replaced by $\widehat{M}$'s, and $C$ is an overall normalization factor.

The superconformal Ward identity also partly determines $M_\text{residual}$ under the assumption that $M_\text{residual}$ has a certain polynomial growth.  For instance, if we require that $M_\text{residual}$ has at most linear growth, as would be the case in a bulk theory of supergravity, then the analytic term is completely fixed in terms of \eqref{Ms} to be \cite{Zhou:2017zaw}
\es{Xcontact}{
M_\text{residual}^\text{SUGRA}
=\frac{1}{2}\left(s+u\sigma^2+t\tau^2-4(t+u)\sigma\tau-4(s+u)\sigma-4(s+t)\tau\right)\,.
}
Thus, the supergravity tree level amplitude takes the form
\es{SUGRATree}{
 M_\text{tree}^\text{SUGRA} = C \left[  \widehat{M}_\text{exchange} + M_\text{residual}^{\rm SUGRA} \right] \,.
 }
For future reference, the linear growth at large $s$, $t$, $u$ is given by\footnote{At large $s$, $t$, $u$, we have  
 $$
  \widehat{M}_\text{exchange}  \approx 
   -\frac{1}{2} \left[ \frac{t^2 + u^2}{s} + \frac{s^2 + t^2}{u} \sigma^2 + \frac{s^2 + u^2}{t} \tau^2 \right] \,.
$$
}
 \es{MSUGRALimit}{
   M_\text{tree}^\text{SUGRA}
   \approx  
     C \left[ \frac{\left(   t u  + s t \sigma + s u \tau \right)^2 }{ s t u} \right]  \,.
 }
The value of the overall coefficient $C$ depends on the normalization of the operators $S_{IJ}$ whose 4-point function we are considering.  It is customary to normalize these operators such that their 2-point function is ${\cal O}(c_T^0)$ at large $c_T$, and then the connected 4-point function scales as $c_T^{-1}$.  In particular, if the normalization of ${\cal O}$ is such that the disconnected piece of the 4-point function is given precisely by \eqref{GDisc}, then the overall coefficient $C$ is fixed to be \cite{Zhou:2017zaw}\footnote{In the notation of  \cite{Zhou:2017zaw}, we have $C =- \lambda_s /  \pi$.} 
\es{lambdas}{
 C  = {32\over \pi^2 c_T} =\frac{3}{2\sqrt{2k}\pi N^{3/2}} + O(N^{-5/2})\,.
}

\subsection{Contribution from higher derivative local terms}

Now suppose the 11D supergravity Lagrangian is deformed by a local term of higher than 2-derivative order. The supersymmetric completion of higher derivative couplings are difficult to write off-shell, but are easily classified through local terms in the flat S-matrix elements of higher momentum powers. In $AdS_4\times S^7$, they give rise to a contribution to the Mellin amplitude that is a polynomial expression in $s, t$, of the form
 \es{GenForm}{
   \left(   t u  + s t \sigma + s u \tau \right)^2  F(s^2 + t^2 + u^2 , s t u) + \ldots \,,
 }
where $F$ is a homogeneous polynomial in $s$, $t$, $u$, determined by the corresponding flat space vertex, and $\cdots$ represents lower degree terms in $s,t$. One can check that the expression \eqref{GenForm} solves the superconformal Ward identity written in Appendix~\ref{SUSYWARD}, after the latter is expanded to leading non-trivial order in large $s$ and $t$.  The number of polynomial solutions to the superconformal Ward identities of degree $p \geq 0$ is thus equal to the number of monomials in $P$ and $Q$, 
 \es{PQDef}{
  P \equiv s^2 + t^2 + u^2 \,, \qquad Q \equiv s t u 
 }
of degree $d_P \geq 0$ in $P$ and degree $d_Q \geq 0$  in $Q$ such that $p \geq 2 d_P + 3 d_Q + 4$.  This number is
 \es{Gotn}{
  n(p)=\Big\lfloor{\frac{6+(p-1)^2}{12}}\Big\rfloor\,.
 }
See the first two lines of Table~\ref{NSol}, where for each degree $p \leq 10$ in $s$, $t$, $u$ we listed the number of local solutions of the Ward identity with that growth at large $s$, $t$, $u$.

\begin{table}[htp]
\begin{center}
\begin{tabular}{c||c|c|c|c|c|c|c|c|c}
degree $ \leq p$  & 3 & 4 & 5 & 6 & 7 & 8 & 9 & 10 & $\cdots$ \\
\hline
\# of solutions & 0 & 1 & 1 & 2 & 3 & 4 & 5 & 7 & $\cdots$ \\[3pt]
 \hline 
 11D vertex &  & $R^4$ & & $D^4 R^4$ & $D^6 R^4$ & $D^8 R^4$ & $D^{10} R^4$ & $D^{12} R^4$ (2 types) & $\cdots$ \\ 
\hline
scaling in M-theory & & ${c_T^{-\frac{5}{3}}}$
 & &  $(0\times){c_T^{-\frac{19}{9}}}$ & ${c_T^{-\frac{7}{3}}}$ & ${c_T^{-\frac{23}{9}}}$ & ${c_T^{-\frac{25}{9}}}$ & ${c_T^{-3}}$, ${c_T^{-3}}\log c_T$ & $\cdots$ \\
 \hline
 spin truncation & & $0$
 & &  $2$ & $3$ & $4$ & $5$ & $6$ & $\cdots$ \\[3pt]
 \end{tabular}
\end{center}
\caption{Number of solutions to the Ward identity of degree $p$ polynomial growth at large $s$, $t$, $u$.  At each order we can always have the solutions from previous orders.  The solution corresponding to $p=1$ is non-analytic;  all other new solutions are purely polynomial in $s$, $t$, $u$ and their number is given by $n(p)$ in \eqref{Gotn}. Spin truncation refers to the maximum spin of operators that receive contributions at this order. In the second to last row, we indicate the order of appearance of the {\it maximal degree} solution in the large $c_T$ expansion of the Mellin amplitude of M-theory on $AdS_4\times S^7$. Note that $D^4R^4$ is expected to be absent in M-theory, while one specific linear combination of the two possible $D^{12}R^4$ terms mixes with the 2-loop logarithmic divergence which is cut off at Planck scale.}\label{NSol}
\end{table}%

Thus, the most general local term in the Mellin amplitude that solves the Ward identity is of the form
 \es{GenSol}{
  M_\text{local} =  C 
  \sum_{p\geq 4} \sum_{k = 1}^{n(p) - n(p-1)}  B_{p, k} M_\text{local}^{(p, k)}  \,,
 }
where $M_\text{local}^{(p, k)}$ is a polynomial solution to the Ward identity of degree $p$, labeled by the index $k$. We left out the overall constant $C$ by convention. A well defined flat space limit would require the coefficients $B_{p,k}$ to scale with the AdS radius $L$ like 
 \es{BScaling}{
B_{p,k} \sim L^{-2(p-1)},~~~~ {\rm as}~L \to \infty \,.
 }

Beyond the leading large $s$, $t$, $u$ asymptotics, the polynomial solutions are quite complicated.    To simplify their form a bit, let us first note that  any function $M(s, t; \sigma, \tau)$ that is crossing invariant can be written as
 \es{MTof}{
  M &=  (1 + \sigma^2 + \tau^2) f_1 + ( s + u \sigma^2 + t \tau^2)f_2  + (s^2 + u^2 \sigma^2 + t^2 \tau^2) f_3 \\
   {}&+  ( \sigma + \tau + \sigma \tau) f_4+  ( t \sigma + u \tau + s \sigma \tau)f_5 + ( t^2 \sigma + u^2 \tau + s^2 \sigma \tau) f_6 \,,
 }
where the $f_i$ are symmetric functions of $s, t, u$, or equivalently functions of $P$ and $Q$ as defined in \eqref{PQDef}.  The first purely polynomial solution to the Ward identity, which is the unique solution of degree $4$ we denoted by $M_\text{local}^{(4, 1)}$ in \eqref{GenSol}, can then be written as
 \es{fPolyn}{
  f_1^{(4, 1)} &=\frac{P^2}{4}  
   + \frac{6}{7} Q - \frac{22}{5} P + \frac{96}{5} \,, \\
  f_2^{(4, 1)} &=  
    Q + 2 P - \frac{736}{35} \,, \\
  f_3^{(4, 1)} &=  
    -\frac{P}{2} + \frac{228}{35} \,, \\
   f_4^{(4, 1)} &=- \frac{104}{7} Q - \frac{40}{7} P + \frac{4672}{35}   \,, \\
  f_5^{(4, 1)} &=  
    2 Q - \frac{18}{7} P - \frac{496}{7} \,, \\
  f_6^{(4, 1)} &=  
    \frac{832}{35} \,.
 } 
In this normalization, the solution $M_\text{local}^{(4, 1)}$ has the asymptotic form \eqref{GenForm} with $F(P, Q) = 1$.  For explicit expressions of all polynomial solutions up to degree 10, see Appendix~\ref{BigPol}.

\subsection{Loop contributions}

While the large $c_T$ expansion of the M-theory Mellin amplitude in $AdS_4\times S^7$ contains local terms that correspond to higher derivative vertices in the flat space limit, there must also be ``loop terms" that are required by unitarity. The loop terms are determined, up to local terms, in terms of lower order terms in the large $c_T$ expansion \cite{Fitzpatrick:2011hu, Fitzpatrick:2011dm, Costa:2012cb, Aharony:2016dwx}.

Unlike the loop terms in the flat spacetime S-matrix, a loop term in the Mellin amplitude involves an infinite series of poles rather than a branch cut in the $s, t, u$ variables. For instance, the supergravity 1-loop Mellin amplitude can be expressed as a sum over poles in $s$ at $s=2\Delta + 2n$, $n=0,1,2,\cdots$, whose residues are polynomials in $t$, together with cross terms related by permutation on $s, t, u$. In the flat space limit, the sum of poles turns into an integral, which is nothing but a representation of the supergravity 1-loop S-matrix in the form of a dispersion relation.

The flat space loop amplitudes can typically be expressed as loop integrals that are UV divergent; the UV divergence can be renormalized by local counter terms up to logarithmic divergences. Similarly, the Mellin loop amplitudes typically involve a divergent sum over poles, that can be regularized by subtracting off polynomials in $s, t$ term by term in the sum, up to logarithmic divergences. The log divergence is physical and is cut off at Planck scale in M-theory, resulting in a $\log c_T$ dependence in the Mellin amplitude. In this paper, we will not compute the M-theory loop Mellin amplitudes explicitly, but illustrate the general structure in a few examples, as follows.

The 1-loop 4-super-graviton amplitude in 11D supergravity has only power divergences that can be renormalized away, resulting in a contribution to the S-matrix element that scales with energy like $\ell_{11}^{18} (\sqrt{s})^{11}$. The 1-loop supergravity contribution to the Mellin amplitude likewise can be written as a convergent sum over double trace poles. It comes with an overall coefficient that scales like $(\ell_{11}/L)^{18}\sim c_T^{-2}$.

In the flat space S-matrix of M-theory, there is a higher momentum order 1-loop amplitude that scales like $\ell_{11}^{24} (\sqrt{s})^{17\over 2}$, whose unitarity cut factorizes into a tree level supergravity amplitude and an $R^4$ vertex. It gives rise to another 1-loop Mellin amplitude that sums up double trace poles, with an overall coefficient that scales like $(\ell_{11}/L)^{24}\sim c_T^{-{8\over 3}}$.

The 2-loop 4-super-graviton amplitude of 11D supergravity has a log divergence of the form \cite{Bern:1998ug} $ (\log\Lambda){7 \over 5\cdot 2^8 \cdot 13!}  \ell_{11}^{18} stu \left[ 438(s^6+t^6+u^6) - 53 s^2t^2u^2 \right] {\cal A}_{\rm SG,tree}$. The cutoff $\Lambda$ is taken to be at Planck scale in M-theory. This gives rise to a local term in the Mellin amplitude of degree 10 in $s,t,u$, whose coefficient scales like $(\ell_{11}/L)^{18} \log(L/\ell_{11}) \sim c_T^{-3}\log c_T$, as indicated in Table~\ref{NSol}.

\subsection{The large radius expansion of the Mellin amplitude of M-theory on $AdS_4\times S^7$}
   
As shown in \cite{Penedones:2010ue,Fitzpatrick:2011hu}, the relation between the large $s$, $t$ limit of the Mellin amplitude $M(s,t)$ and the flat spacetime scattering amplitude ${\cal A}(s,t)$ takes the form 
 \es{FlatLimit}{
\lim_{L\to \infty} (2L)^7 V_7 \, M(L^2 \tilde s, L^2 \tilde t) = \frac{1}{\sqrt{\pi}} \int_0^\infty d \beta\, \beta^{-1/2} e^{-\beta} {\cal A} \left( {2 \beta}\tilde s, {2 \beta} \tilde t \right)  \, ,
 }
where $V_7=\pi^4/3$ is the volume of the unit $S^7$. The amplitude ${\cal A}$ appearing on the RHS is the 11D flat spacetime amplitude of four supergravitons, with momenta restricted to a 4D sub-spacetime, integrated against four supergraviton Kaluza-Klein mode wave functions on the $S^7$, normalized by multiplying with an $S^7$ volume factor so that the $L\to \infty$ limit is finite. Indeed, the scaling in \eqref{BScaling} is such that only the most divergent term in each $M_\text{local}^{(p, k)}$ contributes to the limit on the LHS of \eqref{FlatLimit}. 

More precisely, if we label by $i,j,k,\ell$ the four supergraviton KK modes, then the amplitude ${\cal A}_{ijk\ell}(s,t)$ appearing on the RHS of (\ref{FlatLimit}) is related to the 11D scattering amplitude ${\cal A}^{11D}_{\A\B\gamma\D}(s,t)$ by
\ie\label{aintrel}
{\cal A}_{ijk\ell}(s,t) = \sum_{\A,\B,\gamma,\D} {\cal A}^{11D}_{\A\B\gamma\D}(s,t) \,V_7 \int_{S^7} d^7x \sqrt{g} \Psi_i^\A(x) \Psi_j^\B(x) \Psi_k^\gamma(x) \Psi_\ell^\D(x).
\fe
Here ${\cal A}^{11D}_{\A\B\gamma\D}(s,t)$ is an invariant tensor in the supergraviton polarizations $\A,\B,\gamma,\D$. $\Psi_i^\A(x)$ is the normalized KK mode wave function for the particle $i$ on a unit $S^7$.

Since on the 3D SCFT side we are studying scalar operators transforming as the ${\bf 35}_c$ of $\mathfrak{so}(8)$, the flat space limit of the 4-point function of these operators corresponds to the scattering amplitude ${\cal A}(s,t)$ of the 11D gravitons in their lowest KK modes, with momenta concentrated in a 4-dimensional sub-spacetime and polarization in the transverse directions. After contraction with $\mathfrak{so}(8)$ polarization vectors and rewriting in terms of the $\mathfrak{so}(8)$ invariants $\sigma,\tau$ (after stripping out a factor of $(Y_1 \cdot Y_2)^2 (Y_3 \cdot Y_4)^2$), the scattering amplitude will be denoted by ${\cal A}(s,t;\sigma,\tau)$. Rather than evaluating the integral in (\ref{aintrel}) directly, we can obtain the answer by reducing the tree level amplitude of the lowest KK modes on $AdS_4\times S^7$ to that of the ${\cal N}=8$ gauged supergravity in $AdS_4$ \cite{deWit:1982bul} (see also \cite{Biran:1983iy}, as well as \cite{Duff:1986hr} for a review), whose flat spacetime limit gives the tree amplitude in 4D ungauged ${\cal N}=8$ supergravity \cite{Cremmer:1978ds,Cremmer:1979up,deWit:1977fk}. The details are explained in Appendix~\ref{N8SUGRA}. The result takes the form
 \es{Aflat}{
   &{\cal A}(s,t;\sigma,\tau) =  {\ell_{11}^9} \frac{\left(   t u  + s t \sigma + s u \tau \right)^2}{ stu}
    \Big[  1 + \ell_{11}^6 f_{R^4}(s, t)   + \ell_{11}^9 f_{\rm 1-loop}(s,t)
    +  \ell_{11}^{12} f_{D^6 R^4}(s, t) 
    \\
    &~~~
    +  \ell_{11}^{14} f_{D^8 R^4}(s, t) + \ell_{11}^{15} f_{{\rm 1-loop},R^4}(s,t) +  \ell_{11}^{16} f_{D^{10} R^4}(s, t)
 + \ell_{11}^{18} f_{\rm 2-loop}(s, t)  + \ell_{11}^{18} f_{D^{12}R^4}(s,t)  + \cdots   \Big] \,, 
 } 
with $f_{R^4} = \frac{stu}{3 \cdot 2^7}$ and $f_{D^6 R^4}(s, t, u) = \frac{(stu)^2}{ 15\cdot 2^{15}}$ as given in \eqref{fR4}. $f_{\rm 1-loop}$ and $f_{\rm 2-loop}$ are known 11D supergravity loop amplitudes. The latter comes with a log divergence, whose counter term can be absorbed into $f_{D^{12}R^4}(s,t)$. $ f_{{\rm 1-loop},R^4}(s,t)$ is the 1-loop amplitude, whose unitarity cut involves an $R^4$ vertex and a tree amplitude, as already mentioned; it is given by a known loop integral with only power divergences that can be regularized in the standard way. The coefficients of the local terms $f_{D^8 R^4}$, $f_{D^{10}R^4}$, $f_{D^{12}R^4}$ are not protected by supersymmetry and are unknown. 

At each order in $c_T^{-1}$, the large $s, t$ limit of the Mellin amplitude (at this specific order) is determined by the flat space limit, i.e.~by a corresponding term in the small momentum expansion of ${\cal A}(s,t)$.  As such, the large $c_T$ expansion of the Mellin amplitude is expected to be of the form
\es{Mflat}{
   &M(s,t;\sigma,\tau) =  c_T^{-1} M_{\rm tree}^\text{SUGRA} + c_T^{-{5\over 3}} M_{R^4}   + c_T^{-2} M_{\rm 1-loop}
    + c_T^{-{7\over 3}} M_{D^6 R^4}(s, t) 
    \\
    &~~~ + c_T^{-{23\over 9}} M_{D^8 R^4} + c_T^{-{8\over 3}} M_{{\rm 1-loop},R^4} + c_T^{-{25\over 9}} M_{D^{10} R^4} + c_T^{-3} M_{{\rm 2-loop}} + c_T^{-3} M_{D^{12} R^4} + \cdots .
 } 
While $M_{R^4}$, for instance, is proportional to the unique solution to the superconformal Ward identity of degree 4 in $s,t$, the term $M_{D^6R^4}$ is a linear combination of three independent solutions to the Ward identity, of degree 7, 6, and 4 respectively. 
We must be careful about the interpretation of the loop terms on the RHS. $M_{\rm 1-loop}$ is determined by the tree level supergravity Mellin amplitudes\footnote{To determine the polar part of $M_{\rm 1-loop}$, we need not only the 4-super-graviton amplitude in $AdS_4$, but also the amplitudes involving 2 gravitons and 2 KK modes in $AdS_4$.} up to the ambiguity of a term proportional to $M_{R^4}$. $M_{{\rm 1-loop}, R^4}$ and $M_{\rm 2-loop}$ are subject to similar ambiguities. Note that $c_T^{-3} M_{\rm 2-loop}$ contains a log divergence that is cut off at Planck scale, resulting in a local term proportional to $c_T^{-3} \log(c_T)$ that is of the same degree as $M_{D^{12}R^4}$.

Based on superconformal Ward identities and the flat space limit, a priori one may expect that other terms suppressed by further powers of $({\ell_{11}/L})^2$, such as terms of the form $c_T^{-{17\over 9}} M_{R^4}$ or $c_T^{-{19\over 9}} M_{R^4}$, would be allowed on the RHS of (\ref{Mflat}). As we will see later, such terms are ruled out by comparison with the known CFT data, namely the $1/c_T$ expansion of the OPE coefficient $\lambda_{(B,+)}^2$. At low derivative orders, this can be understood from the supersymmetry protected terms in the bulk effective action as follows. A term suppressed by extra powers of $(\ell_{11}/L)^2$ in comparison to those that survive the flat space limit should come from the reduction of higher-than-4-point effective coupling of the super-graviton on $AdS_4\times S^7$, e.g. terms in the effective action of the schematic form $R^5, R^6$, etc. As explained in \cite{Wang:2015aua}, the $R^5$ type coupling is not compatible with supersymmetry, whereas an $R^6$ coupling should be tied to $D^4 R^4$ by supersymmetry Ward identities, but the latter is absent in the M-theory effective action. This leaves $R^7$, which is tied to $D^6R^4$, and its reduction on $AdS_4\times S^7$ may lead to a contribution to the 4-super-graviton Mellin amplitude that is down by $(\ell_{11}/L)^6 \sim c_T^{-{2\over 3}}$ in comparison to the $R^4$ contribution. This is indeed consistent with the powers of $c_T^{-1}$ appearing in the expansion of $\lambda_{(B,+)}^2$ on the CFT side.\footnote{Beyond order $c_T^{-{7\over 3}}$, however, it is not clear from the bulk why the contributions from, say $c_T^{-{23\over 9}}M_{D^8 R^4}$, to $\lambda_{(B,+)}^2$ should vanish. We will return to this point in Section~\ref{CONCLUSION}.}

Comparing \eqref{FlatLimit}, \eqref{Aflat}, and \eqref{Mflat}, we can determine, up to an overall normalization constant,
\ie\label{mcexp}
M(s,t;\sigma,\tau) =C \left[ \widehat M_{\rm exchange} + M_{\rm residual}^{\rm SUGRA} + B_{4, 1}  M_{\rm local}^{(4,1)}+ ({\rm 1-loop}) + B_{6, 1} M_{\rm local}^{(6,1)} + B_{7, 1} M_{\rm local}^{(7,1)} + \cdots \right]
\fe
where with the normalizations $F(P,Q)=1$ and $F(P,Q)=Q$ for $M^{(4,1)}_\text{local}$ and $M^{(7,1)}_\text{local}$, respectively, we have
\es{GotB}{
  B_{4, 1} \approx  \frac{35}{2^{7}}\, \frac{\ell_{11}^6}{L^6} 
   \,, \qquad
   B_{6, 1} = o\left( \frac{\ell_{11}^{10}}{L^{10}}\right) \,, \qquad B_{7, 1} \approx \frac{9009}{2^{15}} \frac{\ell_{11}^{12}}{L^{12}} 
 }
in the large radius limit.  Using the relation \eqref{LellpRelation} given by the AdS/CFT dictionary, we can write \eqref{GotB} as 
 \es{GotBAgain}{
  B_{4, 1} \approx \frac{70}{ (6\pi c_T k)^{\frac23}} 
  \,, \qquad
   B_{6, 1} = o(c_T^{-10/9}) \,, \qquad
   B_{7, 1} \approx \frac{1001}{2}\left(\frac{6}{\pi^2 c^2_T k^2}\right)^{\frac23} 
   \,.
 }
In the normalization of $S_{IJ}$ in which the disconnected piece of the 4-point function is given precisely by \eqref{GDisc}, the overall coefficient $C$ is given by \eqref{lambdas}, which is exact in $1/c_T$.  This is essentially because the exchange of the stress tensor multiplet only appears in $M_{\rm tree}^\text{SUGRA}$, and hence the coefficient of the latter in the Mellin amplitude is exactly proportional to $c_T^{-1}$. All other terms on the RHS of (\ref{mcexp}) involve exchange of multi-trace operators.

So far, using the known part of the M-theory effective action, we have determined the following terms in the large $c_T$ expansion of the super-graviton Mellin amplitude in $AdS_4\times S^7$: order $c_T^{-1}$ (tree level supergravity), order $c_T^{-{5\over 3}}$ (degree 4 in $s, t$, related to $R^4$ coupling), and the coefficient of the maximal degree 7 polynomial in $s, t$ at order $c_T^{-{7\over 3}}$ (however, we cannot fix the three other coefficients, of degree $6, 5, 4$ polynomials in $c_T^{-{7\over 3}} M_{D^6 R^4}$). In principle, one can fix the non-analytic part of $M_{\rm 1-loop}$ and $M_{{\rm 1-loop}, R^4}$ in terms of the lower order Mellin amplitude (that involves super-gravitons as well as KK modes in $AdS_4$). We also know the order $c_T^{-3} \log c_T$ term that is fixed by the logarithmic divergence of 2-loop amplitude in 11D supergravity. Other coefficients, such as those appearing in $M_{D^8 R^4}$, are entirely unknown due to our ignorance of the higher order terms in the small momentum expansion of the M-theory S-matrix.

In the next section, we show how to relate these coefficients to CFT data, namely the OPE coefficients and scaling dimensions. Thus, if one has an independent way of computing those CFT data, one can reconstruct the corresponding part of the Mellin amplitude.

\section{Comparison with CFT data}
\label{extractcft}

We will now extract CFT data from the tree-level Mellin amplitudes computed above. We will focus on the OPE coefficients squared $a_\mathcal{M}$ of the protected multiplets $\mathcal{M}$ in Table \ref{opemult}, as well as the scaling dimension $\Delta_{A_{(0,j)}}$ of the lowest twist long multiplet with spin $j$. The supergravity contribution to these quantities was computed in \cite{Zhou:2017zaw,Chester:2018lbz}, and by definition is order $c_T^{-1}$. The higher derivative Mellin amplitudes $M^{(p,d)}_\text{local}$ discussed above will contribute starting at order $c_T^{-\frac{7+2p}{9}}$, and then will generically include all subleading powers of $c_T^{-2/9}$ corresponding to powers of $\ell_{11}^2$ in the flat space limit.

As discussed in \cite{Heemskerk:2009pn,Heslop:2017sco,Alday:2014tsa}, a flat space vertex with $2p$ derivatives for $p>1$, which corresponds to an $AdS_4$ Mellin amplitude of maximal degree $p$, contributes to operators with spin $j\leq p-4$. From the list of conformal primaries for $(A,+)_j$ and $(A,2)_j$ in Tables 6 and 7 in \cite{Chester:2014fya}, respectively, we see that these supermultiplets contain a superconformal descendent with spin $j+2$ that is the only operator with these quantum numbers, so these multiplets receives contribution only for $p\geq j+6$. We will now show how to fix the $n(p) - n(p-1)$ coefficients $B_{p,d}$, indexed by $d$, of each degree $p$ tree level term $M^{(p,d)}_\text{local}$ in \eqref{mcexp} by extracting at least $n(p) - n(p-1) $ different pieces of CFT data from these amplitudes, following the algorithm in \cite{Chester:2018lbz}.  

We begin by writing the position space $\mathcal{G}^{(p,d)}$ corresponding to a given $M^{(p,d)}_\text{local}$ as
\es{Aexpansion}{
\mathcal{G}^{(p,d)}(U,V;\sigma,\tau)=\sum_{\mathcal{M}_{\Delta,j}} \left[ a_\mathcal{M}^{(p,d)} \mathfrak{G}_\mathcal{M}(U,V;\sigma,\tau)+ a_{\mathcal{M}}^{(0)} \Delta_\mathcal{M}^{(p,d)}\partial_{\Delta}\mathfrak{G}_{\mathcal{M}}(U,V;\sigma,\tau)\right]_{\Delta^{(0)}_\mathcal{M}} \,,
}
where the subscript ${\Delta^{(0)}_\mathcal{M}} $ denotes that the blocks for the unprotected operators should be evaluated with the leading order scaling dimension. Note that this expression only holds for tree level amplitudes that scale as some fraction of $c_T^{-1}$; for loop terms there would be additional terms. The superblocks $\mathfrak{G}_\mathcal{M}(U,V;\sigma,\tau)$ can be further expanded into $\mathfrak{so}(8)$ structures $Y_{ab}(\sigma,\tau)$ and conformal blocks $G_{\Delta',j'}(U,V)$ as in \eqref{GExpansion}. To compare to the Mellin space amplitude, we will furthermore take the lightcone expansion $U\ll1$ for fixed $V$, so that the conformal blocks can be written as
\es{lightBlocksExp}{
G_{\Delta,j}(U,V)=\sum_{k=0}^\infty U^{\frac{\Delta-j}{2}+k}g_{\Delta,j}^{[k]}(V)\,,
}
where the lightcone blocks $g_{\Delta,j}^{[k]}(V)$ are labeled by the $k+1$-th lowest twist, and are only functions of $V$. For instance, the $k=0$ block in the normalization of \cite{Chester:2018lbz} is
\es{lightconeBlock}{
g_{\Delta,j}^{[0]}(V)&=\frac{\Gamma(j+1/2)}{4^\Delta\sqrt{\pi}j!}(1-V)^j \,{}_2F_1\left(\frac{\Delta+j}{2},\frac{\Delta+j}{2},\Delta+j,1-V\right)\,.\\
}
Note that $g_{\Delta,j}^{[k]}(V)$ goes like $(1-V)^{j-2k}$ in the $V\to 1$ limit.

Putting these ingredients together, we can now expand $\mathcal{G}^{(p,d)}$ to get the final expression
\es{Aexpansion2}{
&\mathcal{G}^{(p,d)}(U,V;\sigma,\tau)= \sum_{a=0}^2 \sum_{b = 0}^a Y_{ab}(\sigma, \tau)  \sum_{\mathcal{M}_{\Delta,j}} \sum_{(\Delta',j')\in\mathcal{M}}\sum_{k=0}^\infty U^{\frac{\Delta'-j'}{2}+k} \\ &\left[  a_\mathcal{M}^{(p,d)} A^\mathcal{M}_{ab \Delta' j'}(\Delta, j) g^{[k]}_{\Delta',j'}(V)
 +a_{\mathcal{M}}^{(0)}   \Delta^{(p,d)}_\mathcal{M}\left[\partial_{\Delta}+\frac{\log U}{2}\right]\left[A^\mathcal{M}_{ab \Delta' j'}(\Delta, j) g^{[k]}_{\Delta',j'}(V)\right]\right]_{\Delta^{(0)}_\mathcal{M}}\,.
}
The utility of the lightcone expansion is that the $U$-dependence corresponds to the twist $\Delta-j$ of a conformal primary, and the $\log U$ term distinguishes between the scaling dimension and the OPE coefficient of that primary. In the Mellin transform \eqref{mellinDef}, one can isolate the $U^{\frac{\Delta'-j'}{2}+k}$ factor by taking the residue of the pole $s=\Delta'-j'+2k$. The $t$-integral can then be performed by summing all the poles, which yields a function of $V$. We can then extract the coefficients of a set of lightcone block using the orthogonality relations for hypergeometric functions \cite{Heemskerk:2009pn}
\es{ortho}{
\delta_{r,r'}&=-\oint_{V=1} \frac{dV}{2\pi i}(1-V)^{r-r'-1} F_r(1-V) F_{1-r'}(1-V)\,,\\
F_r(x)&\equiv {}_2 F_1(r,r,2r,x)\,,
}
where the integration contour is chosen to encircle only the pole $V=1$. For instance, by multiplying $\mathcal{G}^{(p,d)}(U,V;\sigma,\tau)$ with $-(1-V)^{-1-\widetilde j} F_{1-\frac{\Delta'+\widetilde j}{2}}(1-V)$ and then evaluating the residue at $V=1$, we will collect contributions from all terms in $\mathcal{G}^{(p)}(U,V;\sigma,\tau)$ that involve the lightcone blocks $g^{[k]}_{\Delta',j'}(V)$ with $j'=\widetilde j,\widetilde j+2,\dots, \widetilde j+2k$, as well as those involving $\partial_{\Delta'} g^{[k]}_{\Delta',j'}(V)$ with $j'<\widetilde j+2k-1$. Combined with our ability to select the twist $\Delta'-j'$ and $R$-symmetry structure $Y_{ab}(\sigma,\tau)$, as well as our knowledge of how each conformal primary contributes to the superconformal multiplet, this is enough to recursively solve for all $\Delta^{(p)}_\mathcal{M}$ and $a^{(p)}_\mathcal{M}$ for each superconformal multiplet $\mathcal{M}_{\Delta,j}$. 

Recall that there are $n+1$ long multiplets $(A,0)_{n,j,q}$ appearing in the OPE with leading order twist $2n+2$, labeled by $q=0,\cdots,n$. For $n>0$, $\Delta^{(p,d)}_{(A,0)_{n,j,q}}$ extracted from the local term in the Mellin amplitude is actually the average of all $q=0,\dots,n$ operators with the same leading order quantum numbers. To avoid this ambiguity, we will only discuss the $n=0$ case (where the label $q=0$ will be omitted in the notation). To extract the anomalous dimension $\Delta^{(p,d)}_{(A,0)_{0,j}}$, we will also need the leading order OPE coefficient squared $a^{(0)}_{(A,0)_{(0,j)}}$, which we list from \cite{Chester:2018lbz} in Table \ref{Avalues}.

 \begin{table}[htp]
\begin{center}
\begin{tabular}{|l|c|}
\hline
 \multicolumn{1}{|c|}{Spin $j$}  &Leading order OPE coefficient squared $a_{(A,0)_{0,j}}^{(0)}$ \\
  \hline
  0&  $\quad\qquad\qquad\qquad\qquad\qquad \quad\;\;32 / 35\approx0.911$\\
  2 &  $\;\;\quad\qquad\qquad\qquad\qquad\quad\; \,2048 / 693\approx2.955$\\
  4 & $ \;\;\qquad\qquad\qquad\quad\;\,1048576/225225\approx4.656$\\
  6 & $ \;\;\qquad\quad\qquad\quad67108864/10669659\approx6.290$\\
  8 & $ \;\;\qquad\quad\;\;34359738368/4350310965\approx7.899$\\
  10 & $ \;\qquad\,2199023255552/231618204675\approx9.494$\\
  12 & $ 2251799813685248/203176892887605\approx11.083$\\
  \hline
\end{tabular}
\end{center}
\caption{Values of leading order OPE coefficients squared $a_{(A,0)_{0,j}}^{(0)}$ for spin $j$.}\label{Avalues}
\end{table} 

\subsection{Matching the $R^4$ term}
\label{R4extract}

We begin by extracting the CFT data that receives corrections from the degree 4 polynomial Mellin amplitude $M_\text{local}^{(4,1)}$ that corresponds to the $R^4$ term. From the discussion above, the multiplets that receive corrections at this order are $(B,+)$, $(B,2)$, and $(A,0)_{n,0,q}$. Since $\lambda^2_{(B,+)}$ and $\lambda^2_{(B,2)}$ are related by \eqref{1dcrossing}, we will only discuss the former.

For $a^{(4,1)}_{(B,+)}$, we take the $s=2$ pole in the Mellin transform \eqref{mellinDef} of $M^{(4,1)}_\text{local}$ given in \eqref{fPolyn} and find that the $UY_{22}$ coefficient is
\es{Bp1}{
\mathcal{G}^{(4)}\big\vert_{UY_{22}}[V]&=-\frac{8\pi^2}{35}\int\frac{dt}{2\pi i}V^{t/2-1}\csc(\pi t/2)^2\\
&=-\frac{16}{35}\frac{\log V}{1-V}\,,
}
where we closed the contour to include all positive poles in $t$. From the expansion \eqref{Aexpansion2}, we then extract the coefficient of $g^{[0]}_{2,0}(V)$ by integrating against $\frac{16F_0(1-V)}{V-1}=\frac{16}{V-1}$ to find
\es{Bpresult}{
a^{(4,1)}_{(B,+)}=CB_{4,1}\oint_{V=1}\frac{dV}{2\pi i}\frac{16\mathcal{G}^{(5/3)}\big\vert_{UY_{22}}}{V-1}=\frac{256}{35}CB_{4,1}\,,
}
where we used $A_{2220}^{(B,+)}(2,0)=1$ for the superconformal primary. We now compare to the localization result \eqref{exact}, and using the SUGRA normalization \eqref{lambdas} we find that the leading $c_T^{-5/3}$ term in $B_{4,1}$ precisely agrees with the result \eqref{GotBAgain} obtained from the $R^4$ effective coupling in 11D.

We can similarly extract the anomalous dimension at order $c_T^{-{5\over 3}}$ for the lowest $j=0$ $(A,0)$ long multiplet by taking the $s=2$ pole in the $SO(8)_R$ singlet channel and using the leading order OPE coefficient $a^{(0)}_{(A,0)_{0,0}}=\frac{32}{35}$ from Table \ref{Avalues}.   We obtain
\es{long}{
\Delta^{(4,1)}_{(A,0)_{0,0}}=-192 C B_{4,1}=-{71680}\left(\frac{6}{\pi^8k^2}\right)^{1/3}c_T^{-5/3}+O(c_T^{-17/9})\,,
}
where we inputted the value of $CB_{4,1}$ determined above.

\subsection{Higher derivative corrections}
\label{Higherextract}

We now show how to extract CFT data from higher degree Mellin amplitudes $M_\text{local}^{(p,d)}$ in terms of their coefficents $B_{p,d}$ for $p=6,7,8,9,10$, where $d=1$ except for $p=10$ where $d=1,2$. For $p<10$ the leading order in $1/c_T$ contributions can be unambiguously extracted from these terms, as they do not mix with loop contributions. For $p=10$, the $c_T^{-3}$ contribution is affected by the as yet unknown 2-loop term, but there is a $c_T^{-3}\log c_T$ that one could unambiguously extract. For all higher terms, the tree level contribution is indistinguishable from the $2$-loop and higher contributions.

Since $\lambda^2_{(B,+)}$ has already been used to fix $B_{4,1}$ in \eqref{Bpresult}, and $\lambda^2_{(B,2)}$ is related to $\lambda^2_{(B,+)}$ by crossing symmetry, we will use the semi-short $\lambda^2_{(A,2)_j}$ and $\lambda^2_{(A,+)_j}$ as well as the lowest twist unprotected $\Delta_{(A,0)_{0,j}}$ for the allowed spin. These calculations will closely follow the SUGRA calculations in \cite{Chester:2018lbz}, except that we use $M^{(p,d)}_\text{local}$ in Appendix \ref{BigPol}. As such we will only briefly sketch the calculations; for more details see \cite{Chester:2018lbz}.

For $(A,+)_j$, we extract its OPE coefficient using the superconformal descendent  $(j+4,j+2)_{[0040]}$, which has the advantage of being the only conformal primary in $\mathcal{M}$ with these quantum numbers for any $j$. If we had chosen the superconformal primary $(j+2,j)_{[0020]}$, then for $j=2$ this primary would have appeared in both $(A,+)_0$ and $(A,+)_2$. Using the explicit coefficients in Appendix C of \cite{Chester:2014fya} and the formula for $M^{(p,d)}_\text{local}$ in Appendix \ref{BigPol}, we can compute $a^{(p,d)}_{(A,+)_j}$ in terms of $CB_{p,d}$, which we list in Table \ref{resultList}.

\begin{sidewaystable}
\begin {tabular} {| c || c | c | c | c | c | c | c |}
\hline
 {CFT data:}&$M_\text{local}^{(4,1)}$  &$M_\text{local}^{(6,1)}$ & $M_\text{local}^{(7,1)}$ &$M_\text{local}^{(8,1)}$ &$M_\text{local}^{(9,1)}$ &$M_\text{local}^{(10,1)}$ &$M_\text{local}^{(10,2)}$ \\
  \hline
\TBstrut  $a_{(A,+)_0}$&  0& $\frac{16384}{1485}$ & $\frac{950272}{6435}$ & $-\frac{131396796416}{467137125}$& $-\frac{422304284672}{5835588759}$ & $\frac{1363121203815907328}{82825301109705}$ & $\frac{577700480155648}{35496557618445}$ \\
  \hline
 \TBstrut  $a_{(A,+)_2}$& 0& $0$ & $0$ & $\frac{67108864}{557375}$& $\frac{499222839296}{7833219625}$ & $-\frac{452294960405547057152}{33820331286462875}$ & $\frac{10719544062509056}{4831475898066125}$ \\
  \hline
 \TBstrut  $a_{(A,+)_4}$& 0& $0$ & $0$ & $0$& $0$ & $\frac{687194767360}{273854581}$ & $0$ \\
  \hline
\TBstrut  $a_{(A,2)_1}$& 0& $-\frac{533430272}{28875}$ & $-\frac{7141523456}{28875}$ & $\frac{584011250925568}{2096128125}$& $\frac{2720900729798656}{130926670875}$ & $-\frac{275809856661297899241472}{43483283082595125}$ & $-\frac{191793642762885136384}{6211897583227875}$ \\
  \hline
 \TBstrut  $a_{(A,2)_3}$& 0& $0$ & $0$ & $-\frac{2813168548052992}{695269575}$& $-\frac{4188231072338673664}{1954231632045}$ & $\frac{40236691147394137885398532096}{109687451241507854715}$ & $-\frac{18018624516515449274368}{241071321409907373}$ \\
  \hline
 \TBstrut  $a_{(A,2)_5}$&0& $0$ & $0$ & $0$& $0$ & $-\frac{1032968730239572115456}{654551570673}$ & $0$ \\
  \hline
 \TBstrut $\Delta_{(A,0)_{0,0}}$& -192& $\frac{15360}{11}$ & $\frac{192000}{11}$ & $-\frac{18059264}{1521}$& $\frac{12428820480}{2375087}$ & $-\frac{11167368386150400}{46400728913}$ & $-\frac{31945339699200}{5930920237}$ \\
  \hline
 \TBstrut   $\Delta_{(A,0)_{0,2}}$& 0& $-1536$ & $-18432$ & $\frac{509591552}{12675}$& $-\frac{78840397824}{2375087}$ & $-\frac{21440053347492298752}{10254561089773}$ & $\frac{3851013479989248}{112687484503}$ \\
  \hline
 \TBstrut     $\Delta_{(A,0)_{0,4}}$& 0& $0$& $0$ & $-32768$& $\frac{5987205120}{182699}$ & $-\frac{628968771433267200}{788812391521}$ & $-\frac{3801661424271360}{112687484503}$ \\
  \hline
 \TBstrut    $\Delta_{(A,0)_{0,6}}$& 0& $0$& $0$ & $0$& $0$ & $-\frac{176947200}{143}$ & $0$ \\
  \hline
\end{tabular}
\caption{Contributions to the OPE coefficients squared $a$ and anomalous dimensions $\Delta$ of various multiplets appearing in $S\times S$ from {\it local} terms in the Mellin amplitude.  Each polynomial $M_\text{local}^{p,k}$ in the Mellin amplitude contributes to a given quantity in the left column an amount equal to the number in the corresponding entry of the table. 
}\label{resultList}

\end{sidewaystable}

The calculation for $(A,2)_j$ is more subtle, because there is no longer a twist 2 conformal primary that only appears in $(A,2)_j$. We choose the conformal primary $(j+4,j+2)_{[0120]}$, which overlaps with superconformal descendents of $(A,+)_{j\pm1}$. Since we have already computed $a^{(p,d)}_{(A,+)_j}$, we can remove them to find the answers for $a^{(p,d)}_{(A,2)_j}$ as given in Table \ref{resultList}.

For $(A,0)_{0,j}$, since we are considering its anomalous dimension, we only need to worry about mixing with other superconformal descendents of $(A,0)_{0,j'}$ for some other $j'$. If we choose the superconformal primary $(j+2,j)_{[0000]}$, then from Table 8 in \cite{Chester:2014fya} we see that a superconformal descendent of $(A,0)_{0,j}$ mixes with $(A,0)_{0,j+4}$. We can take into account this mixing by computing each $j$ starting from $j=0$, which yields the answers in Table \ref{resultList}.

Note that all the OPE coefficients and anomalous dimensions in Table~\ref{resultList} receive contributions from non-local terms in the Mellin amplitude, such as the tree level amplitude at order $c_T^{-1}$, the 1-loop amplitude at order $c_T^{-2}$, etc.

\section{Discussion}
\label{CONCLUSION}

In this paper, we outlined a strategy to recover the M-theory effective action, i.e.~the small momentum expansion of the flat spacetime S-matrix, from the CFT data of ABJM theory using the large $c_T$ expansion of the Mellin amplitude. We determined certain low order terms in the latter expansion using the OPE coefficient of the $(B,+)$ multiplet, previously computed exactly as a function of $c_T$ via the supersymmetric localization method. The known CFT data are enough for us to recover the correct $R^4$ effective coupling of M-theory, but not enough for a nontrivial check against the next two known coefficients of the M-theory effective action allowed by supersymmetry, namely $D^4 R^4$ (whose coefficient is zero) and $D^6 R^4$. It is plausible that there may be other protected OPE coefficients, say of semi-short multiplets, in the $S\times S$ OPE that could be determined using CFT methods, and tested against the absence of the $D^4R^4$ term and the coefficient of the $D^6R^4$ term in M-theory.

More importantly, our hope is that bootstrap bounds on unprotected OPE coefficients or anomalous dimensions at large $c_T$ could be used to bound the coefficients of higher order terms in the M-theory effective action, such as $D^8 R^4$, $D^{10} R^4$, etc. It has been suggested \cite{Russo:1997mk}, based on naive power counting arguments, that the independent local terms in the M-theory effective action only arise at momentum order $D^{6k} R^4$ for non-negative integer $k$. It is not clear to us why this should be the case beyond $D^6 R^4$, where supersymmetry no longer constrains the moduli dependence of the higher derivative couplings upon toroidal compactifications of M-theory \cite{Wang:2015aua}. Nonetheless, we saw that a certain cancelation in the contribution from local terms in the Mellin amplitude of the form $c_T^{-{23\over 9}} M_{D^8 R^4}$ and $c_T^{-{25\over 9}} M_{D^{10} R^4}$ to the $(B,+)$ OPE coefficient is required, and we do not have an explanation of this from the bulk perspective. This does not imply the absence of $D^8 R^4$ or $D^{10} R^4$ terms in M-theory, however, since the local Mellin amplitudes $M_{D^8 R^4}$ and $M_{D^{10} R^4}$ are not entirely fixed by their flat space limits. An intriguing possibility is that perhaps such terms are absent in the Mellin amplitude altogether (which would imply the absence of $D^8 R^4$ and $D^{10} R^4$ in the flat space limit). It would be extremely interesting to understand if this is the case.

In \cite{Agmon:2017xes}, it was noticed that the $(B, 2)$ (or $(B, +)$) OPE coefficients of ABJ(M) theory, as computed using supersymmetric localization, come close to saturating the numerical bootstrap bounds on these quantities obtained for general ${\cal N} = 8$ SCFTs\@.  Such a bound saturation would imply that one may extract numerically all the CFT data encoded in the $\langle SSSS \rangle$ 4-point function,\footnote{If the numerical bounds are only close to being saturated, then we cannot reconstruct the $\langle SSSS \rangle$ 4-point function, but we can still obtain stringent bounds on the CFT data.} thus allowing us in principle to recover the entire M-theory super-graviton S-matrix using the procedure outlined in this paper.  However, as was pointed out in \cite{Agmon:2017xes}, the values of the $(B, 2)$ OPE coefficients as a function of $1/c_T$ start to depend on $k$ at order $1/c_T^{5/3}$, with the value for the $k=2$ ABJ and ABJM theories being closer to the numerical bound.\footnote{As already mentioned, the OPE coefficients of $k=2$ ABJM and $k=2$ ABJ theories have identical perturbative expansions in $1/c_T$.  See Footnote~\ref{FootnoteNonPert}.}  So it is possible that one of these $k=2$ theories could in fact saturate the bootstrap bound for all values of $c_T$, and the strategy of determining the CFT data numerically and feeding it into the procedure described in this paper could work.  As far as the $k=1$ ABJM theory is concerned, while (at least at large $c_T$) this theory certainly does not saturate the bootstrap bound discussed in \cite{Agmon:2017xes}, it is possible that an improved bootstrap analysis could generate different stronger bounds that apply only to the $k=1$ theory.  For instance, a mixed correlator study of the lowest dimension scalars in the $(B, +)$ $[0020]$ and $[0030]$ multiplets would single out the $k=1$ theory because the $(B, +)$ $[0030]$ multiplet does not exist in the $k=2$ theories.  It would be very interesting to investigate these issues in the future.

So far we have focused entirely on 4-particle S-matrix elements. Our strategy based on the flat space limit of ABJM correlators  allows us, in principle, to recover the M-theory S-matrix elements of $n$ supergravitons, provided that their momenta are aligned within a 4D sub-spacetime of the 11D Minkowskian spacetime. This determines all $n$-point S-matrix elements for $n\leq 5$, but not for $n\geq 6$. To recover the $(n\geq 6)$-point S-matrix elements for general 11D momenta from the Mellin amplitudes of ABJM theory would be much more difficult, as it would require taking a flat space limit of the Mellin amplitudes for operators of large $\mathfrak{so}(8)$ quantum numbers.

It would also be useful to extend the arguments of this paper to other cases of maximally supersymmetric SCFTs with holographic duals, such as $\mathcal{N}=4$ SYM in 4D, which is dual to Type IIB String theory, and the $A_{N-1}$ series of $(2,0)$ theories in 6D, which is dual to M-theory. As mentioned before, none of the CFT data in the stress tensor four point functions in these cases is known analytically beyond $1/c_T$ order, but it is possible that numerical bounds could be translated into bounds on M-theory and String theory. In 6D, the OPE coefficients of certain protected operators that appear in four point function half-BPS multiplets other than the stress tensor are known in an expansion to all orders in $1/c_T$ using the protected 2D chiral algebra. In an upcoming work \cite{Chester:2018dga}, this data will be used to derive the M-theory $R^4$ from 6D CFT, analogous to the 3D derivation in this work.

Lastly, it would be interesting to generalize the construction in this paper to theories with lower amounts of supersymmetry.  In particular, it should be possible to extend the arguments of this paper to the full family of ABJM theories, which have only ${\cal N} = 6$ supersymmetry for CS level $k>2$.  The supersymmetric localization calculations extend to this case too, and one can perform both an expansion in large $N$ at fixed $k$, as we did in this paper, or at large $N$ and fixed $\lambda = N/k$ \cite{Drukker:2010nc}.  The latter expansion would allow us to probe scattering amplitudes in type IIA string theory directly.

\section*{Acknowledgments} 

We thank Ofer Aharony for collaboration at early stages of this work, as well as Nicholas Agia, Nathan Agmon, Igor Klebanov, Eric Perlmutter, Leonardo Rastelli, Yifan Wang, and Alexander Zhiboedov for useful discussions.  SMC and SSP are supported in part by the Simons Foundation Grant No~488651.  SMC is also supported in part by the Bershadsky Family Scholarship in Science or Engineering.  XY is supported by a Simons Investigator Award from the Simons Foundation and by DOE grant DE-FG02-91ER40654.

\appendix

\section{Supersymmetric Ward identity}
\label{SUSYWARD}

In position space, the supersymmetric Ward identity takes the form \cite{Dolan:2004mu}
\es{3D}{
\left(z\partial_z-\frac{1}{2}\alpha\partial_\alpha\right)\mathcal{G}(U,V;\sigma,\tau)\big\vert_{\alpha=z^{-1}}=\left(\bar z\partial_{\bar z}-\frac{1}{2}\alpha\partial_{\alpha}\right)\mathcal{G}(U,V;\sigma,\tau)\big\vert_{\alpha=\bar z^{-1}}=0\,,\\
}
where we defined
\es{zalpha}{
U\equiv z\bar z\,,\qquad V\equiv(1-z)(1-\bar z)\,,\qquad \sigma\equiv\alpha\bar\alpha\,,\qquad\tau\equiv(1-\alpha)(1-\bar\alpha)\,.
}
To implement the Ward identities in Mellin space, we first expand ${\cal G}(U,V;\sigma,\tau)$ into the R-symmetry polynomials $Y_{ab}(\sigma,\tau)$ as
\ie
{\cal G}(U,V;\sigma,\tau) = \sum_{a=0}^2\sum_{b=0}^a Y_{ab}(\sigma,\tau){\cal G}_{ab }(U,V) \,,
\fe
which has Mellin transform \eqref{mellinDef}
\es{mellinWard}{
{M}(s,t;\sigma,\tau) = \sum_{a=0}^2\sum_{b=0}^a Y_{ab}(\sigma,\tau){M}_{ab }(s,t) \,.
}
If we add up the two equations in \eqref{3D}, and expand in powers of $\bar\alpha$, then $z$ and $\bar z$ always appear in the combination $z^m+\bar{z}^m$ for some integer $m$, which can then be turned into rational functions of $U,V$. The resulting equation involves a set of differential operators in $U,V$ acting on ${\cal G}_{ab}(U,V)$, organized in powers of $\bar\alpha$. Finally, we convert the Ward identity to Mellin space by setting
\ie
{\cal G}_{ab }(U,V)\to M_{ab}(s,t),~~~U \partial_U \to \widehat{U\partial_U},~~~ V\partial_V \to \widehat{V\partial_V},~~~ U^mV^n \to \widehat{U^mV^n},
\fe
where the hatted operators act on $M _{ab}(s,t)$ as
\es{3DMellin}{
\widehat{U\partial_U} M _{ab}(s,t)&=\frac s 2  M _{ab}(s,t)\,,\\
\widehat{V\partial_V} M _{ab}(s,t)&=\left[\frac t 2-1\right] M _{ab}(s,t)\,,\\
\widehat{U^mV^n} M _{ab}(s,t)&=M _{ab}(s-2m,t-2n)\left(1-\frac{s}{2}\right)_m^2\left(1-\frac{t}{2}\right)_n^2\left(1-\frac{u}{2}\right)_{-m-n}^2\,,
}
where $u=4-s-t$ and we will have independent constraints on each coefficient in the expansion in powers of $\bar\alpha$.

\section{Polynomial solutions of degree $p\leq10$ }
\label{BigPol}

Here we record the purely polynomial solutions $M_{p,d}$ to the superconformal Ward identity with maximal degree $p$. For $p=6,7,8,9$ we find one new solution for each $p$, while for $p=10$ we find two new solutions. We will write these polynomials in the notation of \eqref{PQDef} and \eqref{MTof}, so that in the large $s,t$ limit they take the form
 \es{GenFormApp}{
&   \left(   t u  + s t \sigma + s u \tau \right)^2  F_{p,d}(P , Q) \,,\\
&P\equiv s^2+t^2+u^2\,,\qquad Q\equiv stu\,,
 }
where they are normalized so that 
\es{largeNorm}{
F_{6,1}=P\,,\quad F_{7,1}=Q\,,\quad F_{8,1}=P^2\,,\quad F_{9,1}=QP\,,\quad F_{10,1}=P^3\,,\quad F_{10,2}=Q^2\,.
}
The full polynomials are then
\es{p6}{
M_\text{local}^{(6,1)}:&\\
f_1^{(6,1)}&=\frac{P^3}{4}-\frac{102 P^2}{11}-\frac{10 P Q}{11}+\frac{1152 P}{11}+\
\frac{608 Q}{77}-\frac{4096}{11}\,,\\
f_2^{(6,1)}&=2 P^2+P Q-\frac{544 P}{11}-16 Q+\frac{21760}{77}\,,\\
f_3^{(6,1)}&=-\frac{P^2}{2}+\frac{180 P}{11}+\frac{40 Q}{11}-\frac{8192}{77}\,,\\
f_4^{(6,1)}&=-\frac{64 P^2}{11}-\frac{208 P Q}{11}+\frac{14912 P}{77}-\frac{8000 \
Q}{77}-\frac{1536}{7}\,,\\
f_5^{(6,1)}&=-\frac{50 P^2}{11}+2 P Q-\frac{11904 P}{77}+\frac{592 \
Q}{11}+\frac{44416}{77}\,,\\
f_6^{(6,1)}&=56 P-\frac{40 Q}{11}-\frac{16320}{77}\,.\\
}

\es{p7}{
M_\text{local}^{(7,1)}:&\\
f_1^{(7,1)}&=\frac{13 P^3}{4}+\frac{P^2 Q}{4}-\frac{1326 P^2}{11}-\frac{2074 P \
Q}{143}+\frac{14976 P}{11}+\frac{36 Q^2}{13}+\frac{15344 \
Q}{143}-\frac{53248}{11}\,,\\
f_2^{(7,1)}&=26 P^2+15 P Q-\frac{7136 P}{11}+Q^2-\frac{2592 Q}{11}+\frac{40960}{11}\,,\\
f_3^{(7,1)}&=-\frac{13 P^2}{2}-\frac{P Q}{2}+\frac{31172 P}{143}+\frac{8500 \
Q}{143}-\frac{204032}{143}\,,\\
f_4^{(7,1)}&=-\frac{11568 P^2}{143}-\frac{37288 P Q}{143}+\frac{394816 \
P}{143}-\frac{296 Q^2}{13}-\frac{97664 Q}{143}-\frac{820736}{143}\,,\\
f_5^{(7,1)}&=-\frac{8460 P^2}{143}+\frac{278 P Q}{13}-\frac{22256 P}{11}+2 \
Q^2+\frac{58368 Q}{143}+\frac{1271936}{143}\,,\\
f_6^{(7,1)}&=\frac{105056 P}{143}+\frac{2720 Q}{143}-\frac{40704}{13}\,.\\
}

\es{p8}{
M_\text{local}^{(8,1)}:&\\
f_1^{(8,1)}&=\frac{P^4}{4}-\frac{118906 P^3}{8775}-\frac{303226 P^2 \
Q}{114075}+\frac{779296 P^2}{2925}+\frac{123799376 P \
Q}{1482975}-\frac{6478336 P}{2925}\\
&+\frac{893008 \
Q^2}{54925}-\frac{776857216 Q}{1482975}+\frac{57749504}{8775}\,,\\
f_2^{(8,1)}&=2 P^3+P^2 Q-\frac{635168 P^2}{8775}-\frac{214816 P \
Q}{7605}+\frac{96849664 P}{114075}+\frac{64736 \
Q^2}{114075}+\frac{7101056 Q}{38025}\\
&-\frac{74051584}{22815}\,,\\
f_3^{(8,1)}&=-\frac{P^3}{2}+\frac{219092 P^2}{8775}+\frac{819392 P \
Q}{114075}-\frac{482840128 P}{1482975}-\frac{16172032 \
Q}{296595}+\frac{1918862848}{1482975}\,,\\
f_4^{(8,1)}&=-\frac{88 P^3}{15}-\frac{344 P^2 Q}{15}+\frac{92261504 \
P^2}{494325}-\frac{807268288 P Q}{1482975}+\frac{669270016 \
P}{1482975}+\frac{20749184 Q^2}{1482975}\\
&+\frac{246585856 Q}{1482975}+\
\frac{3609677824}{1482975}\,,\\
f_5^{(8,1)}&=-\frac{98 P^3}{15}+2 P^2 Q-\frac{11447968 P^2}{32955}+\frac{257584528 \
P Q}{1482975}+\frac{22751488 P}{8775}+\frac{129472 \
Q^2}{114075}\\
&+\frac{638850176 Q}{494325}-\frac{15066962944}{1482975}\,,\\
f_6^{(8,1)}&=\frac{6768 P^2}{65}-\frac{112 P Q}{15}-\frac{883842304 \
P}{1482975}-\frac{116637824 Q}{296595}+\frac{305705984}{164775}\,.\\
}

\es{p9}{
M_\text{local}^{(9,1)}:&\\
f_1^{(9,1)}&=-\frac{4 P^4}{182699}+\frac{P^3 Q}{4}+\frac{256 \
P^3}{182699}-\frac{1470246 P^2 Q}{215917}-\frac{6144 \
P^2}{182699}+\frac{16 P Q^2}{17}+\frac{2169915952 P \
Q}{30876131}\\
&+\frac{65536 P}{182699}+\frac{13084704 \
Q^2}{30876131}-\frac{7754383488 Q}{30876131}-\frac{245760}{182699}\,,\\
f_2^{(9,1)}&=-\frac{32 P^3}{182699}+\frac{365382 P^2 Q}{182699}+\frac{814496 \
P^2}{182699}+P Q^2-\frac{77230848 P Q}{2375087}-\frac{250089728 \
P}{2375087}\\
&-\frac{14173656 Q^2}{2375087}+\frac{246246272 Q}{2375087}+\
\frac{1292496896}{2375087} \,,\\
f_3^{(9,1)}&=\frac{8 P^3}{182699}-\frac{P^2 Q}{2}+\frac{991112 \
P^2}{182699}+\frac{49147600 P Q}{2375087}-\frac{2231011968 \
P}{30876131}+\frac{64 Q^2}{17}\\
&-\frac{3759586080 \
Q}{30876131}+\frac{7327562240}{30876131}\,,\\
f_4^{(9,1)}&=-\frac{90136 P^3}{16609}-\frac{2773792 P^2 \
Q}{182699}+\frac{7526374144 P^2}{30876131}-\frac{456 P \
Q^2}{17}+\frac{21774069376 P Q}{30876131}\\
&-\frac{53524939264 \
P}{30876131}-\frac{14438168832 Q^2}{30876131}+\frac{996369408 \
Q}{165113}-\frac{561678131200}{30876131}\,,\\
f_5^{(9,1)}&=-\frac{9926 P^3}{182699}-\frac{1203696 P^2 Q}{182699}-\frac{175354688 \
P^2}{30876131}+2 P Q^2-\frac{16185059776 P \
Q}{30876131}\\
&+\frac{248339584 P}{182699}+\frac{237662432 \
Q^2}{2375087}-\frac{77421998592 \
Q}{30876131}+\frac{24500873216}{2806921}\,,\\
f_6^{(9,1)}&=\frac{1265488 P^2}{182699}+\frac{20966800 P \
Q}{182699}-\frac{13022172672 P}{30876131}-\frac{64 \
Q^2}{17}+\frac{1028671104 Q}{1816243}\\
&-\frac{57606243328}{30876131}\,.\\
}

\es{p101}{
M_\text{local}^{(10,1)}:&\\
f_1^{(10,1)}&=\frac{P^5}{4}-\frac{15560102282570 \
P^4}{788812391521}-\frac{1352721664990 P^3 \
Q}{46400728913}+\frac{488239952498560 \
P^3}{788812391521}\\
&+\frac{761509603801472 P^2 \
Q}{788812391521}-\frac{7568372894878720 \
P^2}{788812391521}-\frac{36285564992176 P \
Q^2}{788812391521}\\
&-\frac{8944052927693568 P \
Q}{788812391521}+\frac{56120322339061760 \
P}{788812391521}-\frac{93811894726016 \
Q^2}{788812391521}\\
&+\frac{34293760356859904 \
Q}{788812391521}-\frac{157293341281452032}{788812391521}\,,\\
f_2^{(10,1)}&=2 P^4+P^3 Q-\frac{86930417851808 \
P^3}{788812391521}-\frac{193714236162928 P^2 \
Q}{788812391521}+\frac{1555772227023104 \
P^2}{788812391521}\\
&-\frac{4531714954240 P \
Q^2}{46400728913}+\frac{3584920148835072 P \
Q}{788812391521}-\frac{737911375243264 \
P}{46400728913}\\
&+\frac{660575198396864 \
Q^2}{788812391521}-\frac{17697616105197568 \
Q}{788812391521}+\frac{2575160352276480}{46400728913}\,,\\
f_3^{(10,1)}&=-\frac{P^4}{2}+\frac{29459546898780 \
P^3}{788812391521}+\frac{2793360500552 P^2 \
Q}{46400728913}-\frac{100360584961920 \
P^2}{71710217411}\\
&-\frac{1858335910889184 P \
Q}{788812391521}+\frac{13493643403236352 \
P}{788812391521}-\frac{15338759568448 \
Q^2}{41516441659}\\
&+\frac{12456326461020928 \
Q}{788812391521}-\frac{53235484975194112}{788812391521}\,,\\
f_4^{(10,1)}&=-\frac{112 P^4}{19}-\frac{512 P^3 Q}{19}+\frac{560077576316736 \
P^3}{788812391521}+\frac{206913777028736 P^2 \
Q}{788812391521}\\
&-\frac{1708367914422272 \
P^2}{71710217411}+\frac{2132575731403584 P \
Q^2}{788812391521}-\frac{53535047508136448 P \
Q}{788812391521}\\
&+\frac{15759894134800384 \
P}{71710217411}+\frac{3255946003526144 \
Q^2}{71710217411}-\frac{54552662836060160 \
Q}{112687484503}\\
&+\frac{937350835787825152}{788812391521}\,,\\
f_5^{(10,1)}&=-\frac{162 P^4}{19}+2 P^3 Q-\frac{73978710764208 \
P^3}{112687484503}+\frac{781824468030880 P^2 \
Q}{788812391521}\\
&+\frac{7762246710240640 \
P^2}{788812391521}-\frac{9063429908480 P \
Q^2}{46400728913}+\frac{45210666149624576 P \
Q}{788812391521}\\
&-\frac{157261767221821440 \
P}{788812391521}-\frac{7746117391574656 \
Q^2}{788812391521}+\frac{132057277767493632 \
Q}{788812391521}\\
&-\frac{30032651544567808}{71710217411}\,,\\
f_6^{(10,1)}&=\frac{54328 P^3}{323}-\frac{216 P^2 Q}{19}-\frac{1729256379496320 \
P^2}{788812391521}-\frac{538093588640640 P \
Q}{41516441659}\\
&+\frac{39075972258892288 \
P}{788812391521}+\frac{15338759568448 \
Q^2}{41516441659}-\frac{31559299392227840 \
Q}{788812391521}\\
&+\frac{75001766595411968}{788812391521}\,.\\
}

\es{p102}{
M_\text{local}^{(10,2)}:&\\
f_1^{(10,2)}&=\frac{299520 P^4}{112687484503}-\frac{195812761 P^3 \
Q}{26514702236}-\frac{23003136 P^3}{112687484503}+\frac{P^2 \
Q^2}{4}+\frac{28123710360 P^2 Q}{112687484503}\\
&+\frac{705429504 \
P^2}{112687484503}-\frac{111982834408 P \
Q^2}{112687484503}+\frac{741648981488 P \
Q}{112687484503}-\frac{11777605632 P}{112687484503}\\
&+\frac{90 \
Q^3}{19}+\frac{284961517968 Q^2}{112687484503}-\frac{7688398979328 \
Q}{112687484503}+\frac{59624128512}{112687484503}\,,\\
f_2^{(10,2)}&=\frac{2396160 P^3}{112687484503}-\frac{350338726 P^2 \
Q}{5930920237}+\frac{521837758752 \
P^2}{112687484503}+\frac{13061538357 P \
Q^2}{6628675559}\\
&-\frac{78313654944 P \
Q}{112687484503}-\frac{12715684780800 \
P}{112687484503}+Q^3-\frac{3295800890632 \
Q^2}{112687484503}\\
&-\frac{3955386769280 \
Q}{112687484503}+\frac{69834972106752}{112687484503}\,,\\
f_3^{(10,2)}&=-\frac{599040 P^3}{112687484503}+\frac{195812761 P^2 Q}{13257351118}+\
\frac{47804861328 P^2}{10244316773}-\frac{P \
Q^2}{2}+\frac{2089851982888 P Q}{112687484503}\\
&-\frac{6420168418176 \
P}{112687484503}+\frac{2447753248668 \
Q^2}{112687484503}-\frac{14510922184416 \
Q}{112687484503}+\frac{16952863051776}{112687484503}\,,\\
f_4^{(10,2)}&=-\frac{525817052976 P^3}{112687484503}-\frac{2126292211816 P^2 \
Q}{112687484503}+\frac{3829309249536 \
P^2}{10244316773}-\frac{3084100585440 P \
Q^2}{112687484503}\\
&+\frac{158246717908608 P \
Q}{112687484503}-\frac{101933809566720 P}{10244316773}-\frac{584 \
Q^3}{19}+\frac{21726894668416 \
Q^2}{10244316773}\\
&-\frac{2789434673572352 \
Q}{112687484503}+\frac{461843423772672}{5930920237} \,,\\
f_5^{(10,2)}&=-\frac{3406879532 P^3}{112687484503}+\frac{7280563630 P^2 \
Q}{112687484503}-\frac{234903259776 \
P^2}{112687484503}-\frac{44350210808 P \
Q^2}{6628675559}\\
&-\frac{1631675597600 P \
Q}{112687484503}+\frac{278557348028672 P}{112687484503}+2 \
Q^3-\frac{80859508899296 Q^2}{112687484503}\\
&+\frac{1258912624870784 \
Q}{112687484503}-\frac{403649143517184}{10244316773} \,,\\
f_6^{(10,2)}&=\frac{624467978240 P^2}{112687484503}+\frac{2192783584032 P \
Q}{112687484503}-\frac{78555860196864 \
P}{112687484503}+\frac{14514678629152 \
Q^2}{112687484503}\\
&-\frac{294378521704192 \
Q}{112687484503}+\frac{1144844596080640}{112687484503}\,.\\
}

\section{Scattering amplitude in 4D ${\cal N} = 8$ supergravity}
\label{N8SUGRA}

In this Appendix we show how to obtain the flat space limit of the tree-level scattering amplitude of the ${\bf 35}_c$ scalars of ${\cal N} = 8$ gauged supergravity, thus deriving the leading term in \eqref{Aflat}.  In the flat space limit, the tree ${\cal N} = 8$ gauged supergravity amplitude reduces to the tree amplitude in ungauged ${\cal N} = 8$ supergravity, which we now review.

The 4D ${\cal N} = 8$ gravity multiplet consists of 128 bosonic and 128 fermionic massless states that can be conveniently represented as anti-symmetric tensors of the $SU(8)$ R-symmetry as follows:  the helicity $h = +2$ and $h=-2$ states of the graviton can be represented as $SU(8)$ singlets $h^+$ and $h^- = h^{ABCDEFGH}$;  the helicity $h = +3/2$ and $h = -3/2$ states of the gravitino can be represented as $\psi^A$ and $\psi^{ABCDEFG}$;  the helicity $h = +1$ and $h = -1$ states of the gravi-photon can be represented as $v^{AB}$ and $v^{ABCDEF}$;  the helicity $h = +1/2$ and $h=-1/2$ states of the gravi-photino can be represented as $\chi^{ABC}$ and $\chi^{ABCDE}$;  and the scalars, of helicity $h=0$, can be represented as $S^{ABCD}$.  Here $A = 1, \ldots 8$ are $SU(8)$ fundamental indices.

The 4-point scattering amplitude of any four particles from the gravity multiplet can be succinctly described by first introducing auxiliary Grassmann variables $\eta_A$ and grouping all the particles of the gravity multiplet into an ${\cal N} = 8$ superfield (see for example \cite{Elvang:2015rqa})
 \es{PhiDef}{
  \Phi = h^+ + \eta_A \psi^A - \frac 12 \eta_A \eta_B v^{AB} - \frac{1}{6} \eta_A \eta_B \eta_C \psi^{ABC} + \frac{1}{4!} \eta_A \eta_B \eta_C \eta_D S^{ABCD} 
   + \cdots  \,.
 }
The expression for $\Phi$ is designed such that one can extract a state of a given helicity by taking derivatives with respect to the auxiliary Grassmann variables $\eta_A$.  For the 70 scalars, we have
 \es{GetS}{
  S^{ABCD} = \frac{\partial}{\partial \eta^A}  \frac{\partial}{\partial \eta^B}  \frac{\partial}{\partial \eta^C}  \frac{\partial}{\partial \eta^D} \Phi \,.
 }  
The tree-level 4-point scattering amplitude in supergravity can then be written as (see \cite{Elvang:2015rqa})\footnote{For the scattering amplitudes corresponding to higher derivative interactions in 4D, see \cite{Bianchi:2008pu,Elvang:2010jv, Freedman:2011uc}.}
 \es{ASG}{
  {\cal A}_\text{tree, SG}(s, t; \eta_i) = \frac{1}{256} \prod_{A=1}^8 \left(\sum_{i, j = 1}^4 \langle i j\rangle \eta_{i A} \eta_{jA} \right)  \frac{[34]^4}{\langle 12 \rangle^4} \frac{1}{stu} \,, 
 }
Here, $\eta_{iA}$, $i = 1, \ldots 4$, are the auxiliary polarization variables associated with the $i$th particle.  The scattering amplitude of 4 scalars can be extracted by acting with derivatives on \eqref{ASG}:
 \es{ScalarAmplitude}{
  {\cal A}_\text{tree, SG} (SSSS )^{A_1 \cdots D_4}(s, t)
   = \partial_1^{A_1B_1C_1D_1} \partial_2^{A_2B_2C_2D_2}
    \partial_3^{A_3B_3C_3D_3}
    \partial_4^{A_4B_4C_4D_4} {\cal A}_\text{tree, SG}(s,t; \eta_i) \,,
 }
where $\partial_i^{ABCD} \equiv \frac{\partial}{\partial \eta_{iA}} \frac{\partial}{\partial \eta_{iB}} \frac{\partial}{\partial \eta_{iC}} \frac{\partial}{\partial \eta_{iD}}$.

To obtain the flat space limit of the scattering amplitude of the ${\bf 35}_c$ scalars in {\em gauged} supergravity, we should identify which of the 70 scalars $S^{ABCD}$ of ungauged supergravity correspond to the ${\bf 35}_c$ ones.  To do so, note that the $SO(8)$ R-symmetry in $AdS_4$ is embedded into the $SU(8)$ flat space R-symmetry in such a way that the supercharges, transforming in the ${\bf 8}$ of $SU(8)$, should also transform as the ${\bf 8}_v$ of $SO(8)$ according to the convention we use in this paper.  The 70 $S^{ABCD}$ scalars transform then as an irreducible representation of $SU(8)$, namely the ${\bf 70}$, which decomposes as ${\bf 35}_s \oplus {\bf 35}_c$ under $SO(8)$---the ${\bf 35}_s$ and ${\bf 35}_c$ can be identified with self-dual and anti-self-dual rank-4 anti-symmetric tensors, respectively.

To connect this discussion to our notation, we should convert between the representation of the ${\bf 35}_c$ as a rank-4 anti-self-dual tensor of the ${\bf 8}_v$ and its representation as a rank-2 symmetric traceless tensor of the ${\bf 8}_c$ that we have been using.  The conversion is realized through a tensor $E^{IJ}{}_{ABCD}$, which is symmetric traceless in the ${\bf 8}_c$ indices $I, J$ and anti-symmetric in the ${\bf 8}_v$ indices obeying the anti-self-duality condition
 \es{asd}{
  E^{IJ}{}_{ABCD} = -\frac{1}{24} \epsilon_{ABCD}{}^{A'B'C'D'} E^{IJ}{}_{A'B'C'D'} \,.
 }
Here, $\epsilon$ is the totally anti-symmetric tensor defined such that $\epsilon^{12345678} = 1$, and all indices are raised and lowered  with the Kronecker symbol.  

To obtain $E^{IJ}{}_{ABCD}$, one can start with the Clebsch-Gordan coefficients $E^I{}_{aA}$ for obtaining an $SO(8)$ singlet out of the product ${\bf 8}_v \otimes {\bf 8}_c \otimes {\bf 8}_s$:  the coefficients $E^I{}_{aA}$ have the property that for any three quantities $u_I$, $v^a$, and $w^A$ transforming as ${\bf 8}_v$, ${\bf 8}_s$, and ${\bf 8}_c$, respectively, the product $u_I v^a w^A E^I{}_{aA}$ is an $SO(8)$ singlet.  As is well-known, the $E^I{}_{aA}$ can be identified with the coefficients in the multiplication table of the generators $e_\alpha$ ($\alpha = 1, \ldots 8$) of the octonion algebra:   $e_\alpha \cdot e_\beta = E^\gamma{}_{\beta \alpha} e^\gamma$, where $e_1 = 1$ and $e_\alpha \cdot e_\alpha = 1$ for any given $\alpha$.  Explicit formulas for the $E^I{}_{aA}$ are given in (A.12) of \cite{Pestun:2007rz}.  From the $E^I{}_{aA}$, we can construct
 \es{EDef1}{
  E^{IJ}{}_{AB} = E^{[I}{}_{a[A} E^{J]a}{}_{B]}  \,,
 }
which is a tensor that converts between the adjoint representation of $SO(8)$ written as either an anti-symmetric tensor of the ${\bf 8}_v$ or as an anti-symmetric tensor of the ${\bf 8}_c$.  Then, using $E^{IJ}{}_{AB}$, we can further construct our desired tensor   
 \es{EDef}{
  E^{IJ}{}_{ABCD} = E^{IK}{}_{AB} E^{JK}{}_{CD}
   + E^{JK}{}_{AB} E^{IK}{}_{CD} - \frac 14 \delta^{IJ} E^{KL}{}_{AB} E^{KL}{}_{CD} \,,
 }
which has all the properties we required.  

From any anti-self-dual anti-symmetric tensor $T^{ABCD}$ we can obtain a symmetric traceless tensor $E^{IJ}{}_{ABCD} T^{ABCD}$, which can be further contracted with the null polarizations $Y^I$ to obtain a quadratic function of $Y$:
 \es{SY}{
  T(Y) = Y_I Y_J E^{IJ}{}_{ABCD} T^{ABCD} \,.
 }
Using this procedure for the amplitude \eqref{ScalarAmplitude}, we can extract 
  \es{ScalarAmplitudeAgain}{
    {\cal A}_\text{tree, SG} (SSSS )(s, t; Y_i) = 
     \left( \prod_{i=1}^4 Y_{iI} Y_{iJ} E^{IJ}{}_{A_iB_iC_iD_i} \right)  {\cal A}_\text{tree, SG} (SSSS )^{A_1 \cdots D_4}(s, t) \,.
  }
Due to the $SO(8)$ R-symmetry, this expression can be written as $ (Y_1 \cdot Y_2)^2 (Y_3 \cdot Y_4)^2 $ times a function of the $SO(8)$ invariants $\sigma$ and $\tau$ introduced in \eqref{uvsigmatauDefs}.  To uncover this form, it is easier to set $Y_i$ to some particular values, for instance
 \es{YParticular}{
  Y_i = \begin{pmatrix} \frac{1 - \vec{y}_i^2}{2} & \vec{y} & i \frac{1 + \vec{y}_i^2}{2}  \end{pmatrix} \,,
 }
for some 6-vectors $\vec{y}_i$ that we can further take to be 
 \es{yChoice}{
  \vec{y}_1 &= \begin{pmatrix} 1 & 0 & 0 & 0 & 0 & 0\end{pmatrix} \,, \\
  \vec{y}_2 &= \begin{pmatrix} \infty & 0 & 0 & 0 & 0 & 0\end{pmatrix} \,, \\
  \vec{y}_3 &= \begin{pmatrix} 0 & 0 & 0 & 0 & 0 & 0\end{pmatrix} \,, \\
  \vec{y}_4 &= \begin{pmatrix} x & y & 0 & 0 & 0 & 0\end{pmatrix} \,, 
 }
for some parameters $x$ and $y$.  Plugging these expressions in \eqref{ScalarAmplitudeAgain} one finds that
 \es{ScAmpAgain}{
  \frac{ {\cal A}_\text{tree, SG} (SSSS )(s, t; Y_i)}{ (Y_1 \cdot Y_2)^2 (Y_3 \cdot Y_4)^2}
   &= \frac{1120}{stu} \langle 34 \rangle^4 [34]^4  \Biggl[ 
     1  - 4 x A + 4(1-x) B 
        + 2 (3x^2 + y^2) A^2 \\
        &\hspace{-1in}{}+\left( 4 x (x-1) + 12 y^2 \right) AB
         + \left( 7 (x-1)^2 + 2 y^2 \right) B^2
          - 4 x (x^2 + y^2) A^3 \\
         &\hspace{-1in} {}+4  \left(x^2 (x-1) + (x-3) y^2 \right) A^2 B 
          + 4 \left(x(x-1)^2 + (2 + x) y^2  \right) A B^2 \\
         &\hspace{-1in}  {}+4 (1-x) \left( (x-1)^2 + y^2) \right) B^3
           - 4 (x^2 + y^2) \left( x (x-1) + y^2 \right) A^3 B \\
         &\hspace{-1in} {}+ (x^2 + y^2)^2 A^4   
          + 2 \left(y^2 + 3 \left( y^2 + x (x-1) \right)^2 \right) A^2 B^2 \\
          &\hspace{-1in} {}-4 \left( (x-1)^2 + y^2 \right) \left( x (x-1) + y^2 \right) A B^3 
          + \left( (x-1)^2 + y^2 \right)^2 B^4 
    \Biggr] \,,
 }
where
 \es{ABDef}{
  A \equiv \frac{\langle 13 \rangle \langle 24 \rangle}{\langle 12 \rangle \langle 34 \rangle}  \,, \qquad
  B \equiv \frac{\langle 14 \rangle \langle 23 \rangle}{\langle 12 \rangle \langle 34 \rangle}  \,.
 }
Making use of the $SO(8)$ symmetry, the $x$ and $y$ dependence can be rewritten in terms of $\sigma$ and $\tau$ through
 \es{xyTost}{
  x = \frac{1 + \sigma - \tau}{2} \,, \qquad
   y^2 = \frac{2 \sigma (1 + \tau) - \sigma^2 - (1 - \tau)^2}{4} \,.
 }

Using that 
 \es{stuDef}{
  s &= (p_3 + p_4)^2 =  -\langle 34 \rangle [34] \,, \qquad
   t = (p_2 + p_3)^2 = -\langle 23 \rangle [23] \,, \qquad \\
    u &= (p_2 + p_4)^2 = -\langle 24 \rangle [24] \,,
 }
as well as the relations 
 \es{ConsRelations}{
  \langle 12 \rangle [24] = - \langle 13 \rangle [34] \,, \qquad
   \langle 12 \rangle [23] = \langle 14 \rangle [34]
 }
that follow from momentum conservation, it can be shown that 
 \es{GotAB}{
  A = -\frac{u}{s} \,, \qquad B = \frac{t}{s} \,.
 }
Plugging \eqref{xyTost} and \eqref{GotAB} into \eqref{ScAmpAgain} and using that $s + t + u = 0$, it can be shown that \eqref{ScAmpAgain} can be rewritten as
 \es{ScalarAmpFinal}{
    \frac{{\cal A}_\text{tree, SG} (SSSS )(s, t; Y_i) }{ (Y_1 \cdot Y_2)^2 (Y_3 \cdot Y_4)^2}
     = 1120 
      \frac{\left(   t u  + s t \sigma + s u \tau \right)^2 }{ s t u}  \,.
 }
Up to an overall constant, we have thus derived the form of the first term in \eqref{Aflat}.

\bibliographystyle{ssg}
\bibliography{R4}

\end{document}